\documentclass{svjour3}                     \smartqed  \usepackage{graphicx}
\usepackage{amsmath}
\usepackage{amssymb}
\usepackage{psfrag}
\usepackage{pstricks}
\usepackage[nospace,noadjust]{cite}
\RequirePackage{pstricks}
\RequirePackage{pst-node}

\newcommand{\fig}[4]{}
\newcommand{\redcolour}{gray}
\newcommand{\bluecolour}{black}
\newcommand{\true}{\texttt{True}}
\newcommand{\false}{\texttt{False}}
\renewcommand{\emptyset}{\varnothing}

\begin{document}

\title{Connectivity Graphs of Uncertainty Regions\thanks{A preliminary extended abstract summarizing parts of this paper appears in~\cite{cef-cgur-10}.}
}

\titlerunning{Connectivity Graphs of Uncertainty Regions}

\author{Erin Chambers$^1$\and Alejandro Erickson$^2$\and S\'andor P.\ Fekete$^3$\and Jonathan Lenchner$^4$\and Jeff Sember$^5$\and Venkatesh Srinivasan$^6$\and Ulrike Stege$^6$\and Svetlana Stolpner$^7$\and\\ Christophe Weibel$^8$\and Sue Whitesides$^6$}
\institute{$^1$~Department of Mathematics and Computer Science, St.~Louis University, MO, USA, echambe5@slu.edu\and $^2$~School of Engineering and Computing Sciences, Durham University, alejandro.erickson@gmail.com\and $^3$~Department of Computer Science, Braunschweig University of Technology, Germany, s.fekete@u-bs.de\and $^4$~IBM Thomas J.~Watson Research Center, Yorktown Heights, NY, USA, lenchner@us.ibm.com\and $^5$~Department of Computer Science, University of British Columbia, BC, Canada, jpsember@cs.ubc.ca
\and $^6$~Department of Computer Science, University of Victoria, BC, Canada, \{venkat, stege, sue\}@cs.uvic.ca \and$^7$~svetlana.stolpner@gmail.com \and $^8$~christophe.weibel@gmail.com}
\date{Received: date / Accepted: date}

\maketitle

\begin{abstract}
\makeatletter{}We study problems of connectivity under uncertainty, where the precise location of
input points is not fixed in advance. We distinguish
two fundamental scenarios under which uncertainty arises.
In the favorable {\em Best-Case Uncertainty} (BU), each input point can be chosen
from a given set, such that the resulting point set has best possible objective value. 
In the unfavorable {\em Worst-Case Uncertainty} (WU), the
input set has worst possible objective
value among all possible choices, e.g., due to imprecise data.  

We consider these notions of uncertainty for the bottleneck spanning tree problem,
giving rise to the following
{\em Best-Case Connectivity with Uncertainty} (BCU) problem: 
Given a family of geometric regions, choose one point per region, such that the longest length of 
an associated geometric spanning tree is minimized.
We show that this problem is NP-hard even for very simple scenarios in which the regions are line
segments or squares. 
On the other hand, we give an exact solution for the case in which 
there are $n$ regions, each consisting of a point and  $k$ regions, each consisting of a line segment, for fixed $k$.  We then give approximation algorithms for cases
where the regions are either all line segments or all unit discs. We
also provide approximation methods for the corresponding {\em Worst-Case Connectivity
with Uncertainty} (WCU) problem:
Given a set of uncertainty regions, find the minimal distance $r$ such that for
any choice of points, one per region, there is a spanning tree among the points
with edge length at most $r$.

\end{abstract}

\makeatletter{}\section{Introduction}\label{sec:intro}

Finding an optimally connected substructure in a network is one of the fundamental combinatorial optimization problems in network design. The standard problem of minimizing the total edge cost in the network amounts to finding a minimum spanning tree, which can be computed by straightforward greedy methods. A closely related problem that has gained in importance in the context of wireless networking is to consider the ``bottleneck'' problem of minimizing the length of the longest edge. 
A solution to this problem allows one to choose a set of lowest power, equi-power routers to be placed at nodes, such that a message can be relayed between any two nodes.
However, the situation changes when the location of devices becomes part of the problem: How should each location be chosen from a given neighborhood, such that the solution to the resulting bottleneck connectivity problem is minimized? The neighborhoods can be the result of imprecise input data, or simply arise from a geometric range of possible locations; depending on the scenario, the choice of locations can be optimistic (i.e., best case) or adversarial (i.e., worst case).

Let $U$, $|U|=n$, denote a family of uncertainty regions, e.g., a family of disks, squares, line segments or pairs of points.  For each uncertainty region $u_i \in U$, $1 \leq i \leq n$, one point $p_i$ is to be chosen inside this region $u_i$.  Let $P$ be the set of points chosen. For some value $\alpha \in \mathbb{R}$, we define the {\em connectivity graph} $G_{\alpha}=(V,E)$ of $P$ with respect to $\alpha$ as follows: $V=P$ and $E = \{ (p_i, p_j)\in P\times P, \|p_i-p_j\|_2 \leq 2\alpha\}$.  Thus, the graph connects a pair of points with an edge whenever closed disks of radius $\alpha$ centered at these points intersect. We can now formally define the main problem, {\em Best-Case Connectivity with Uncertainty} (BCU), that we study in this paper.\\

\noindent
{\bf The BCU Problem.}
\label{problem2}
Given a set $U=\{u_1,\ldots, u_n\}$ of $n$ uncertainty regions, find the minimum value $\alpha$ for which there exists a choice of point set $P=\{p_1,\ldots, p_n\}, p_i \in u_i$, such that the connectivity graph $G_{\alpha}$ of $P$ is connected.

\medskip
\noindent We further study a closely related problem, {\em Worst-Case  Connectivity with Uncertainty} (WCU):\\

\noindent
{\bf The WCU Problem.}
Given a set $U=\{u_1,\ldots, u_n\}$ of $n$ uncertainty regions, find the minimum value $\alpha$,
such that for {\em any} choice of point set $P=\{p_1,\ldots, p_n\}, p_i \in u_i$ the connectivity graph $G_{\alpha}$ of $P$ is connected.

\subsection{Related Work.}
If the $n$ uncertainty regions are points (in other words, there is no uncertainty),
then finding the minimum $\alpha$ for which the connectivity graph is connected amounts to finding a
minimum Euclidean Bottleneck Spanning Tree (MBST) on the points.  Because minimum spanning trees
(MSTs) are also MBSTs, these can be found in time $O(n\log{n})$.

Closely related to our BCU and WCU problems is the  well-studied family of {\rm range assignment} problems.
In these problems, the disks centered at each point can be of different radii,
and the goal is to minimize the total power consumption under the constraint
that the network satisfies certain structural properties like connectivity,
strong connectivity, or a particular broadcast property.
  Most of the work on these problems has considered point sets rather than
uncertainty regions (see \cite{cps-opapr-00,lp-eaasb-02,lp-ptasb-05,SCG06,fuchs-06}).
  Thus our work provides an early exploration of connectivity problems, arising in the context of wireless networks, for points lying in nontrivial uncertainty regions.

The minimum spanning tree problem (MST) has been studied in the setting of
uncertainty regions. Yang {\em et al.} \cite{Yang07} showed that the problem of
computing a spanning tree that minimizes the total edge length is NP-hard if
the uncertainty regions are non-overlapping unit disks or rectangles. They also
give a polynomial-time approximation scheme (PTAS) for the case in which the
uncertainty regions are unit disks; this is notably different from our problem,
which does not admit a PTAS, unless P=NP. Further approximation results
for minimization and maximization versions of the MST with uncertainty were provided by 
Dorrigiv et al.~\cite{dfh+-mmwst-12}.
Another optimization problem with
neighborhoods that have received attention is the Traveling Salesman Problem;
e.g.\ see \cite{ah-aagcp-94,mm-aagtn-95,gl-faatn-99,dm-aatnp-01,delm-tsp-03,bgk+-tnvs-05,m-ptnfr-07}.
The bottleneck version of TSP is known to be NP-hard \cite[p.~212]{Garey_Johnson79}. 
A $2$-approximation has been known since 1984 \cite{parker_rardin84}.

Other work on geometric optimization with uncertainty region has been phrased in setting with imprecise data or with neighborhoods. 
In the context of shortest paths, see~\cite{pk-aspas-10,dmmw-rsprm-14,dmm-mspip-15}.
For a more general treatment and discussion of problems such as convex hulls, see L{\"o}ffler and van Kreveld~\cite{lv-lschi-10},
who also considered the size of bounding boxs, diameters and related problems in~\cite{lv-lbbsm-10}.
A discrete variant in $d$-dimensional space was considered by Ding and Xu~\cite{dx-scccp-11}, who studied
the problem if picking one representative each from a family of finite sets, such that the resulting set has
a small enclosing hypersphere.   

Another angle is to consider topological changes under uncertainty: when does the structure of an optimal solution change
when the input data is perturbed? Abellanas et al.~\cite{ahr-stdt-99} study this structure with respect to the largest 
perturbation of a set of planar points that keeps the Delaunay triangulation unchanged. Conversely, problems of determining
a necessary perturbation in order to achieve a desired change have also been studied; e.g., see Arkin et al.\cite{adk+-ct-11}
for deciding whether a given set of neighborhoods has a convex stabber, which amounts to deciding whether a given set
can be moved into a convex position. For further discussions of related problems, see the excellent exposition by
L{\"o}ffler and van Kreveld~\cite{lv-lschi-10}.

\subsection{Our Main Results.}
After showing that several variants of BCU are NP-hard (some even to approximate), we give exact and approximation algorithms for certain variants.
Given the geometric nature of our problems, we use the Euclidean measure of distance.
Our main results are as follows:
\begin{enumerate}
\item
We show that BCU is NP-hard even in the simple cases in which the uncertainty regions are point pairs and vertical line segments, respectively.
Our proof technique also works when the regions are all squares.
We further show that it is NP-hard to approximate BCU within a factor less than $\sqrt{5}/2$ when the uncertainty regions are pairs of points.
See Section~\ref{sec:hardness}.
\item
We present an exact algorithm for BCU when the instance consists of $n$ fixed points and $k$ line segments. The algorithm is polynomial in $n$ for constant $k$. The output of this algorithm is correct up to precision $\delta$, $\delta > 0$. See Section~\ref{sec:exactAlgorithm}.
\item
For uncertainty regions that are all unit disks, we give a simple constant additive approximation algorithm for BCU. A slight modification of this algorithm gives a constant multiplicative approximation in case the disks are \emph{non-overlapping}. See Section~\ref{sec:constant_factor_and_additive_approx}.
\item
We provide approximation results for the WCU. In particular, we establish methods with additive and multiplicative performance guarantees. See Section~\ref{sec:wcu}.
\end{enumerate}

\makeatletter{}\section{Hardness Results for BCU} \label{sec:hardness}

We prove hardness results for three variants of the BCU problem. Our first main result shows NP-hardness when the uncertainty regions are point pairs (Theorem~\ref{theorem1}). Interestingly,  this result also implies a hardness of approximation result for the case of point pairs  (Theorem~\ref{theorem2}), and NP-hardness when the uncertainty regions are line segments  (Theorem~\ref{theorem3}). Our second main result shows NP-hardness when the uncertainty regions are unit squares (Theorem~\ref{theorem4}).  We assume, in all cases, that the uncertainty regions are non-overlapping.  All of our reductions are from Planar 3-SAT -- in other words 3-SAT with the added condition that the input formula can be represented as a planar graph.

\subsection{BCU when uncertainty regions are point pairs.}
\label{sec:point-pairs}

We consider the BCU problem for uncertainty regions of vertically aligned pairs of points, unit distance apart with integer coordinates.
We study the decision version of the BCU problem for $\alpha = 1$, {\em i.e.}, we want to decide if $G_{\alpha} = G_1$ is connected for some choice of points, one for each uncertainty pair.
\begin{theorem}\label{theorem1} It is NP-hard to find an exact solution to the BCU problem for the case in which the regions of uncertainty are point pairs
that have a vertical distance of length one.
\end{theorem}

\begin{proof}  We show this problem is NP-hard, using a reduction from the following formulation of Planar 3-SAT. Let $\Phi = (X, C)$ be an instance of 3-SAT, with variables $X= \{x_1, \ldots, x_n\}$ and clauses $C = \{c_1, \ldots, c_m\}$.  Each clause consists of exactly three literals, each a variable or its negation.  For such an instance, we define a formula graph $H(\Phi)$ as follows: $H(\Phi) = (V, E)$ with vertex set $V =X\cup C$ and edge set $E=E_1\cup E_2$, such that $E_1 =\{(x_i,x_{i+1}) | i<n\}$, and $E_2=\{(x_i,c_j) | c_j \mbox{ contains } x_i \mbox{ or } \overline{x_i}\}$. A Planar 3-SAT instance is one whose corresponding formula graph $H(\Phi)$ is planar. In the Planar 3-SAT problem, our goal is to determine whether a given Planar 3-SAT instance $\Phi$ is satisfiable. This problem is known to be NP-complete \cite{Garey_Johnson79,Lichtenstein}.

Our reduction makes use of the fact that, given a Planar 3-SAT instance $\Phi$ with formula graph $H(\Phi)$, this graph has a planar layout on an $O(n+m) \times O(n+m)$ grid~\cite{Duchet,Rosenstiehl}.  Further, in this layout, the vertices (variables and clauses) can be drawn as horizontal line segments and edges as vertical line segments.  Henceforth, we equate the formula graph $H(\Phi)$ with the planar layout we have described above.

To reduce from Planar 3-SAT to  BCU when the uncertainty regions are pairs of points, we design various gadgets.  Specifically, given a layout of a Planar 3-SAT instance using line segments as described above, we replace each horizontal line segment corresponding to a variable by a variable gadget, each horizontal line segment corresponding to a clause by a clause gadget, and each vertical line segment corresponding to an edge in $E_1$ by a variable-variable connector and each one corresponding to an edge in $E_2$ by a variable-clause connector.  Below we will argue that there exists a choice of point in each of these uncertainty pairs such that the connectivity graph for $\alpha = 1$, $G_1$, is connected if and only if the corresponding Planar 3-SAT instance is satisfiable.

\subsubsection*{Overview of the gadgets}
We  present the main ideas behind the clause gadget, variable gadget and connector gadgets mentioned above.

A clause gadget is designed so that it contains three ``gates'', one for each of the literals in the clause. The gate for each
literal is either on the top or the bottom of the clause gadget, depending on whether the literal appears below or above the
clause in the planar grid layout of $H(\Phi)$. For the connectivity subgraph corresponding to the
clause to be connected to the rest of the graph in $G_1$, at least one of these three gates must be open. This corresponds
to setting the literal to \true\ in the clause. This, in turn, ensures that the clause is satisfied. 
The role of a variable gadget is to choose and propagate a truth value
for the variable to all the clauses containing it in a consistent
manner. The variable gadget contains three types of constructs. Type
$I$ and type $II$ constructs help link the variable to all the clauses
that contain it and are either above or below it. We have one such
type $I$-type $II$ pair for every occurrence of the variable in a
clause. A construct of type $III$ is used to ensure that the subgraph
corresponding to the variable gadget can be connected if and only if
the truth assignment to the variable in all the copies of type
$I$-type $II$ pairs are the same.  We also construct $G_1$ so that it
does not join subgraphs arising from parts of a variable
gadget. 
The variable and clause gadgets are linked to each other using two
types of connectors. The clause-variable connector replaces an edge of
$H(\Phi)$ between a clause vertex and a variable vertex in such a way
that it connects the corresponding gadgets (in $G_1$) if and only if
the truth value of the variable is consistent with its occurrence (as
a literal) in the clause.  The variable-variable connector replaces an
edge of $H(\Phi)$ between two variable vertices. It connects one
variable gadget to another variable gadget in $G_1$ irrespective of
the choice of truth values for each of
them. 
Having given the overview of the reduction, we now provide a detailed proof by first describing the different  gadgets in detail and then  arguing the correctness of our reduction.

\subsubsection*{The clause gadget}

Figure~\ref{schema_clause_points} depicts a schema that describes the functioning of a clause gadget.  Each gate in the schema represents the entry of a connection to a literal, with an open gate representing a contribution of \true.  If all gates are closed, then, as suggested by the schema, it is possible that the connectivity graph of the clause gadget is connected, but it is isolated from the rest of the graph.  Also, as the schema suggests,  if gates to two literals $x_i$ and $x_j$ are both open, then connections are created between the clause gadget and the gadgets for the literals, but no connection via the clause gadget is made between the variable gadgets.

\makeatletter{}\begin{figure}[h]
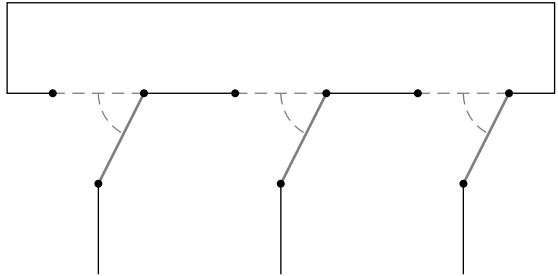

\begin{center}
 \psset{unit=0.6cm,arrows=-,shortput=nab,linewidth=0.5pt,arrowsize=2pt 5,labelsep=5.5pt}
 \pspicture(-1,0.5)(11,7)
 \psline(1,0)(1,2)
 \psline(5,0)(5,2)
 \psline(9,0)(9,2)
 \psset{linecolor=gray,linewidth=1pt}
 \psline(2,4)(1,2)
 \psline(6,4)(5,2)
 \psline(10,4)(9,2)
 \psset{linecolor=gray,linestyle=dashed,linewidth=0.5pt}
 \psline(2,4)(0,4)
 \psline(6,4)(4,4)
 \psline(10,4)(8,4)
 \psarc(2,4){1}{180}{240}
 \psarc(6,4){1}{180}{240}
 \psarc(10,4){1}{180}{240}
 \psset{linecolor=black,linestyle=solid,linewidth=0.5pt}
 \psline(2,4)(4,4)
 \psline(6,4)(8,4)
 \psline(10,4)(11,4)(11,6)(-1,6)(-1,4)(0,4)
 \psdots(2,4)(0,4)(6,4)(4,4)(10,4)(8,4)(1,2)(5,2)(9,2)
  \endpspicture
 \vspace*{10pt}
 \caption{Schema of the clause gadget. When a gray gate is open the
   entire clause gadget can be connected to the rest of the graph.}
\label{schema_clause_points}
\end{center}
\end{figure}

The shape of the clause gadget can be adapted to meet the requirements of the clause vertex it represents in the planar layout of $H(\Phi)$ (e.g., the length of the horizontal line segment representing a particular clause gadget in the planar layout of $H( \Phi )$ by line segments;  however many of the horizontal segments representing literals contained in the clause lie below the segment representing the clause and however many above).  Figure~\ref{clause_gadget_points} shows an example of a clause gadget where, in the representation of $H (\Phi )$, the clause was represented by a horizontal segment connected to one horizontal variable segment lying above, and two horizontal variable segments lying below the segment for the clause. The clause gadget is flexible, as its size can be adjusted by adding more uncertainty pairs to the sequence between two connections to the variable gadgets, and to the sequence between connections to variable gadgets and the left and right sides of the gadget.  Furthermore, in a straightforward manner we can modify the clause gadget to move the connection to a particular variable gadget vertically by one unit without moving the entire clause gadget; e.g., the uncertainty pairs in the gray box in Figure~\ref{clause_gadget_points} can be moved up by one unit.

Consider the three uncertainty pairs with white and gray points in Figure~\ref{clause_gadget_points}.  If all three of the white points are chosen, then it is easy to see that, while points can be chosen so that the connectivity subgraph arising from pairs in the clause gadget can be made connected, no such subgraph can be connected to the rest of $G_1$ for any choice of points in the remaining uncertainty pairs.    If a gray point is chosen from a white-gray pair, then the dashed edges shown incident to the pair do not belong to  $G_1$.

\makeatletter{}\begin{figure}[h]
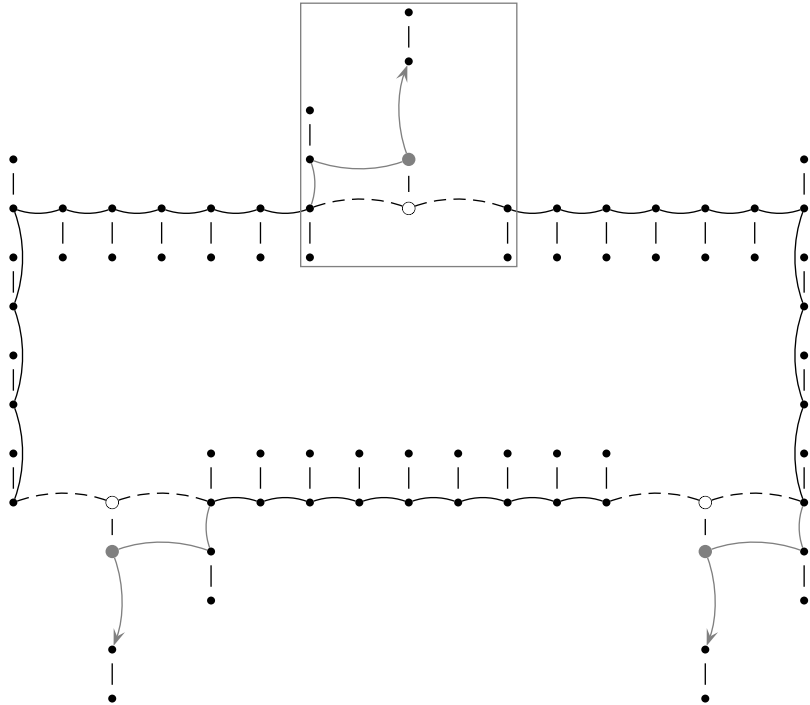

\begin{center}
 \psset{unit=1.3cm,arrows=-,shortput=nab,linewidth=0.5pt,arrowsize=2pt 5,labelsep=5.5pt}
 \pspicture(0,2)(8,8.5)
  \dotnode(1,1.5){a0}
 \dotnode(4,8.5){a1}
 \dotnode(7,1.5){a2}
  \dotnode[linewidth=1.5pt,linecolor=gray](1,3){a3}
 \dotnode(2,2.5){a4}
 \dotnode(2,3.5){a5}
 \dotnode(2.5,3.5){aa5}
 \dotnode(3,3.5){a6}
 \dotnode(3.5,3.5){aa6}
 \dotnode(4,4){a7}
 \dotnode(3,7){a8}
 \dotnode(4.5,3.5){aa7}
 \dotnode(5,3.5){a9}
 \dotnode(5.5,3.5){aa8}
 \dotnode(6,3.5){a10}
 \dotnode[linewidth=1.5pt,linecolor=gray](7,3){a11}
 \dotnode(8,2.5){a12}
 \dotnode(8,3.5){a13}
 \dotnode(8,4.5){a14}
 \dotnode(8,5.5){a15}
 \dotnode(8,6.5){a16}
 \dotnode(7.5,6.5){aa16}
 \dotnode(7,6.5){a17}
 \dotnode(6.5,6.5){aa17}
 \dotnode(6,6.5){a18}
 \dotnode(5.5,6.5){aa18}
 \dotnode(5,6.5){a19}
 \dotnode[linewidth=1.5pt,linecolor=black,dotstyle=o](4,6.5){a20}
 \dotnode(3,6.5){a21}
 \dotnode(2.5,6.5){aa21}
 \dotnode(2,6.5){a22}
 \dotnode(1.5,6.5){aa22}
 \dotnode(1,6.5){a23}
 \dotnode(0.5,6.5){aa23}
 \dotnode(0,6.5){a24}
 \dotnode(0,5.5){a25}
 \dotnode(0,4.5){a26}
 \dotnode(0,3.5){a27}

 \dotnode(1,2){b0}
 \dotnode(4,8){b1}
 \dotnode(7,2){b2}
 \dotnode[linewidth=1.5pt,linecolor=black,dotstyle=o](1,3.5){b3}
 \dotnode(2,3){b4}
 \dotnode(2,4){b5}
 \dotnode(2.5,4){bb5}
 \dotnode(3,4){b6}
 \dotnode(3.5,4){bb6}
 \dotnode(4,3.5){b7}
 \dotnode(4.5,4){bb7}
 \dotnode(3,7.5){b8}
 \dotnode(5.5,4){bb8}
 \dotnode(5,4){b9}
 \dotnode(6,4){b10}
 \dotnode[linewidth=1.5pt,linecolor=black,dotstyle=o](7,3.5){b11}
 \dotnode(8,3){b12}
 \dotnode(8,4){b13}
 \dotnode(8,5){b14}
 \dotnode(8,6){b15}
 \dotnode(8,7){b16}
 \dotnode(7.5,6){bb16}
 \dotnode(7,6){b17}
 \dotnode(6.5,6){bb17}
 \dotnode(6,6){b18}
 \dotnode(5.5,6){bb18}
 \dotnode(5,6){b19}
 \dotnode[linewidth=1.5pt,linecolor=gray](4,7){b20}
 \dotnode(3,6){b21}
 \dotnode(2.5,6){bb21}
 \dotnode(2,6){b22}
 \dotnode(1.5,6){bb22}
 \dotnode(1,6){b23}
 \dotnode(0.5,6){bb23}
 \dotnode(0,7){b24}
 \dotnode(0,6){b25}
 \dotnode(0,5){b26}
 \dotnode(0,4){b27}

 \psset{linecolor=black,nodesep=.1,linestyle=dashed}
 \ncline{a0}{b0}
 \ncline{a1}{b1}
 \ncline{a2}{b2}
 \ncline{a3}{b3}
 \ncline{a4}{b4}
 \ncline{a5}{b5}
 \ncline{a6}{b6}
 \ncline{a7}{b7}
 \ncline{a8}{b8}
 \ncline{a9}{b9}
 \ncline{a10}{b10}
 \ncline{a11}{b11}
 \ncline{a12}{b12}
 \ncline{a13}{b13}
 \ncline{a14}{b14}
 \ncline{a15}{b15}
 \ncline{a16}{b16}
 \ncline{a17}{b17}
 \ncline{a18}{b18}
 \ncline{a19}{b19}
 \ncline{a20}{b20}
 \ncline{a21}{b21}
 \ncline{a22}{b22}
 \ncline{a23}{b23}
 \ncline{a24}{b24}
 \ncline{a25}{b25}
 \ncline{a26}{b26}
 \ncline{a27}{b27}
 \ncline{aa5}{bb5}
 \ncline{aa6}{bb6}
 \ncline{aa7}{bb7}
 \ncline{aa8}{bb8}
 \ncline{aa16}{bb16}
 \ncline{aa17}{bb17}
 \ncline{aa18}{bb18}
 \ncline{aa21}{bb21}
 \ncline{aa22}{bb22}
 \ncline{aa23}{bb23}

 \psset{linecolor=gray,linestyle=solid,nodesep=0pt,arcangle=20,arrows=-}
 \ncarc{->}{a3}{b0}
 \ncarc{a3}{b4}
 \ncarc{b4}{a5}
 \ncarc{->}{b20}{b1}
 \ncarc{b20}{a8}
 \ncarc{a8}{a21}
 \ncarc{->}{a11}{b2}
 \ncarc{a11}{b12}
 \ncarc{b12}{a13}

 \psset{linecolor=black,linestyle=dashed}
 \ncarc{a27}{b3}
 \ncarc{b3}{a5}
 \ncarc{a10}{b11}
 \ncarc{b11}{a13}
 \ncarc{a20}{a19}
 \ncarc{a21}{a20}

  \psset{linecolor=black,linestyle=solid}
 \ncarc{a5}{aa5}
 \ncarc{aa5}{a6}
 \ncarc{a6}{aa6}
 \ncarc{aa6}{b7}
 \ncarc{b7}{aa7}
 \ncarc{aa7}{a9}
 \ncarc{a9}{aa8}
 \ncarc{aa8}{a10}
 \ncarc{a13}{a14}
 \ncarc{a14}{a15}
 \ncarc{a15}{a16}
 \ncarc{a16}{aa16}
 \ncarc{aa16}{a17}
 \ncarc{a17}{aa17}
 \ncarc{aa17}{a18}
  \ncarc{a18}{aa18}
 \ncarc{aa18}{a19}
 \ncarc{a21}{aa21}
 \ncarc{aa21}{a22}
  \ncarc{a22}{aa22}
 \ncarc{aa22}{a23}
 \ncarc{a23}{aa23}
 \ncarc{aa23}{a24}
 \ncarc{a24}{a25}
 \ncarc{a25}{a26}
 \ncarc{a26}{a27}

 \psframe[linearc=.25,linecolor=gray](2.9,5.9)(5.1, 8.6)
 
 \endpspicture
\end{center}
 \vspace*{10pt}
 \caption{An example clause gadget.  As shown, aside from vertical
   sequences of uncertainty pairs leading to variable gadgets, there
   are no other uncertainty pairs in the vicinity of the clause
   gadget.  The graph $G_1$ can be connected only if at least one of
   the gray points is chosen; that is, the attached literal is set to
   \true. The gate inside the gray box can be moved up one unit to
   meet variable-clause connectors at different heights, if
   necessary.}
\label{clause_gadget_points}
\end{figure}

Each of the three white-gray uncertainty pairs is connected to a variable gadget by a sequence of vertical uncertainty pairs.  The choice of a gray point, shown in the schema as an open gate, is intended to mean that the literal (a variable or its negation) connecting to this open gate contributes a \true\ to the clause.  Note that the clause gadget never connects two variable gadgets.

Next, we outline how variable gadgets transmit truth values and how consistency of truth assignments is assured.

\subsubsection*{The variable gadget}
An example of the variable gadget is shown in
Figure~\ref{vble_gadget_points}.  Let the uncertainty pair at the
extreme left of the variable gadget be the \emph{reference pair} for
this variable.  We adopt the interpretation that the choice of black
point in the reference pair means a setting of \true\ to the variable
and the choice of gray point means a setting of \false\ to the
variable.

\makeatletter{}\begin{figure}[h]
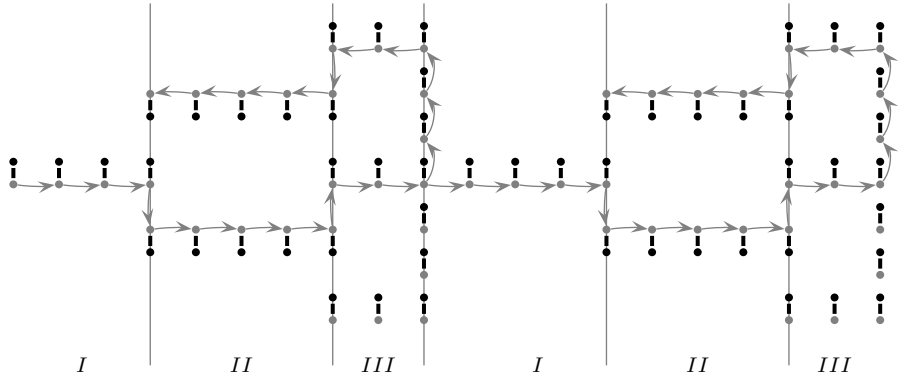

\begin{center}
 \psset{unit=0.3cm,arrows=-,shortput=nab,linewidth=0.5pt,arrowsize=2pt 5,labelsep=5.5pt}
 \pspicture(0,-2)(38,14)

 \psline[linecolor=gray](6,-2)(6,14)
 \psline[linecolor=gray](14,-2)(14,14)
 \psline[linecolor=gray](18,-2)(18,14)
 \psline[linecolor=gray](26,-2)(26,14)
 \psline[linecolor=gray](34,-2)(34,14)

\rput(3,-2){$I$}
\rput(10,-2){$II$}
\rput(16,-2){$III$}
\rput(23,-2){$I$}
\rput(30,-2){$II$}
\rput(36,-2){$III$}

 \psset{linecolor=gray}
 \dotnode(0,6){a0}
 \dotnode(2,6){a1}
 \dotnode(4,6){a2}
 \dotnode(6,6){a3}
 \dotnode(6,4){a4}
 \dotnode(8,4){a5}
 \dotnode(10,4){a6}
 \dotnode(12,4){a7}
 \dotnode(14,4){a8}
 \dotnode(6,10){a9}
 \dotnode(8,10){a10}
 \dotnode(10,10){a11}
 \dotnode(12,10){a12}
 \dotnode(14,10){a13}
 \dotnode(14,0){a14}
 \dotnode(16,0){a15}
 \dotnode(18,0){a16}
 \dotnode(18,2){a17}
 \dotnode(18,4){a18}
 \dotnode(18,6){a19}
 \dotnode(18,8){a20}
 \dotnode(18,10){a21}
 \dotnode(18,12){a22}
 \dotnode(16,12){a23}
 \dotnode(14,12){a24}
 \dotnode(14,6){a25}
 \dotnode(16,6){a26}

 \dotnode(20,6){c0}
 \dotnode(22,6){c1}
 \dotnode(24,6){c2}
 \dotnode(26,6){c3}
 \dotnode(26,4){c4}
 \dotnode(28,4){c5}
 \dotnode(30,4){c6}
 \dotnode(32,4){c7}
 \dotnode(34,4){c8}
 \dotnode(26,10){c9}
 \dotnode(28,10){c10}
 \dotnode(30,10){c11}
 \dotnode(32,10){c12}
 \dotnode(34,10){c13}
 \dotnode(34,0){c14}
 \dotnode(36,0){c15}
 \dotnode(38,0){c16}
 \dotnode(38,2){c17}
 \dotnode(38,4){c18}
 \dotnode(38,6){c19}
 \dotnode(38,8){c20}
 \dotnode(38,10){c21}
 \dotnode(38,12){c22}
 \dotnode(36,12){c23}
 \dotnode(34,12){c24}
 \dotnode(34,6){c25}
 \dotnode(36,6){c26}

 \psset{linecolor=black,dotstyle=*}

 \dotnode(0,7){b0}
 \dotnode(2,7){b1}
 \dotnode(4,7){b2}
 \dotnode(6,7){b3}
 \dotnode(6,3){b4}
 \dotnode(8,3){b5}
 \dotnode(10,3){b6}
 \dotnode(12,3){b7}
 \dotnode(14,3){b8}
 \dotnode(6,9){b9}
 \dotnode(8,9){b10}
 \dotnode(10,9){b11}
 \dotnode(12,9){b12}
 \dotnode(14,9){b13}
 \dotnode(14,1){b14}
 \dotnode(16,1){b15}
 \dotnode(18,1){b16}
 \dotnode(18,3){b17}
 \dotnode(18,5){b18}
 \dotnode(18,7){b19}
 \dotnode(18,9){b20}
 \dotnode(18,11){b21}
 \dotnode(18,13){b22}
 \dotnode(16,13){b23}
 \dotnode(14,13){b24}
 \dotnode(14,7){b25}
 \dotnode(16,7){b26}

 \dotnode(20,7){d0}
 \dotnode(22,7){d1}
 \dotnode(24,7){d2}
 \dotnode(26,7){d3}
 \dotnode(26,3){d4}
 \dotnode(28,3){d5}
 \dotnode(30,3){d6}
 \dotnode(32,3){d7}
 \dotnode(34,3){d8}
 \dotnode(26,9){d9}
 \dotnode(28,9){d10}
 \dotnode(30,9){d11}
 \dotnode(32,9){d12}
 \dotnode(34,9){d13}
 \dotnode(34,1){d14}
 \dotnode(36,1){d15}
 \dotnode(38,1){d16}
 \dotnode(38,3){d17}
 \dotnode(38,5){d18}
 \dotnode(38,7){d19}
 \dotnode(38,9){d20}
 \dotnode(38,11){d21}
 \dotnode(38,13){d22}
 \dotnode(36,13){d23}
 \dotnode(34,13){d24}
 \dotnode(34,7){d25}
 \dotnode(36,7){d26}

 \psset{linecolor=black,nodesep=.1,linestyle=dashed,linewidth=1.5pt}
 \ncline{a0}{b0}
 \ncline{a1}{b1}
 \ncline{a2}{b2}
 \ncline{a3}{b3}
 \ncline{a4}{b4}
 \ncline{a5}{b5}
 \ncline{a6}{b6}
 \ncline{a7}{b7}
 \ncline{a8}{b8}
 \ncline{a9}{b9}
 \ncline{a10}{b10}
 \ncline{a11}{b11}
 \ncline{a12}{b12}
 \ncline{a13}{b13}
 \ncline{a14}{b14}
 \ncline{a15}{b15}
 \ncline{a16}{b16}
 \ncline{a17}{b17}
 \ncline{a18}{b18}
 \ncline{a19}{b19}
 \ncline{a20}{b20}
 \ncline{a21}{b21}
 \ncline{a22}{b22}
 \ncline{a23}{b23}
 \ncline{a24}{b24}
 \ncline{a25}{b25}
 \ncline{a26}{b26}

 \ncline{c0}{d0}
 \ncline{c1}{d1}
 \ncline{c2}{d2}
 \ncline{c3}{d3}
 \ncline{c4}{d4}
 \ncline{c5}{d5}
 \ncline{c6}{d6}
 \ncline{c7}{d7}
 \ncline{c8}{d8}
 \ncline{c9}{d9}
 \ncline{c10}{d10}
 \ncline{c11}{d11}
 \ncline{c12}{d12}
 \ncline{c13}{d13}
 \ncline{c14}{d14}
 \ncline{c15}{d15}
 \ncline{c16}{d16}
 \ncline{c17}{d17}
 \ncline{c18}{d18}
 \ncline{c19}{d19}
 \ncline{c20}{d20}
 \ncline{c21}{d21}
 \ncline{c22}{d22}
 \ncline{c23}{d23}
 \ncline{c24}{d24}
 \ncline{c25}{d25}
 \ncline{c26}{d26}

 \psset{linecolor=gray,nodesep=.1pt,linestyle=solid,linewidth=0.5pt,arrows=->}
\psset{arcangle=-8}
 \ncarc{a0}{a1}
 \ncarc{a1}{a2}
 \ncarc{a2}{a3}
 \ncarc{a3}{a4}
\psset{arcangle=8}
 \ncarc{a4}{a5}
 \ncarc{a5}{a6}
 \ncarc{a6}{a7}
 \ncarc{a7}{a8}
 \ncarc{a8}{a25}
\psset{arcangle=-8}
 \ncarc{a25}{a26}
 \ncarc{a26}{a19}
\psset{arcangle=-40}
 \ncarc{a19}{a20}
 \ncarc{a20}{a21}
 \ncarc{a21}{a22}
\psset{arcangle=8}
 \ncarc{a22}{a23}
 \ncarc{a23}{a24}
 \ncarc{a24}{a13}
\psset{arcangle=-8}
 \ncarc{a13}{a12}
 \ncarc{a12}{a11}
 \ncarc{a11}{a10}
 \ncarc{a10}{a9}
 \ncarc{a19}{c0}
 \ncarc{c0}{c1}
 \ncarc{c1}{c2}
 \ncarc{c2}{c3}
 \ncarc{c3}{c4}
\psset{arcangle=8}
 \ncarc{c4}{c5}
 \ncarc{c5}{c6}
 \ncarc{c6}{c7}
 \ncarc{c7}{c8}
 \ncarc{c8}{c25}
\psset{arcangle=-8}
 \ncarc{c25}{c26}
 \ncarc{c26}{c19}
\psset{arcangle=-40}
 \ncarc{c19}{c20}
 \ncarc{c20}{c21}
 \ncarc{c21}{c22}
\psset{arcangle=8}
 \ncarc{c22}{c23}
 \ncarc{c23}{c24}
 \ncarc{c24}{c13}
\psset{arcangle=-8}
 \ncarc{c13}{c12}
 \ncarc{c12}{c11}
 \ncarc{c11}{c10}
 \ncarc{c10}{c9}
 \endpspicture
\caption{An example variable gadget.}
\label{vble_gadget_points}
\end{center}
\end{figure} 

In order for the constructs of type $I$, $II$, and $III$ to function
as described in the Overview of the gadgets, above, we require that
the following two properties hold: for any choice of points in all
uncertainty pairs, no cycle of $G_1$ contains a pair of
non-consecutive vertices in variable gadgets; and, the subgraph of
$G_1$ restricted to a variable gadget can be connected if and only if
the truth values in type $I$ and $II$ constructs are consistent with
the reference pair (as these are the parts of the gadget that
propagate truth values to clause gadgets).  Together, these properties
ensure that for $G_1$ to be connected, the variable gadgets must be
internally connected, and, in turn, each variable gadget must
propagate consistent truth values to the clause gadgets that are
connected to it. We prove these two properties now.

Recall that in $G_1$ the clause gadget connects to the literals that
are satisfied (only) by opening gates, and that no two open gates are
connected through the clause gadget.  As such, there cannot be a path
in $G_1$ joining two (possibly indistinct) variable gadgets through a
clause gadget.  In addition, the edges $E_1$ form a path of variable
vertices (horizontal lines) in $H(\Phi)$, but not a cycle, so that
combined with the above, no two non-consecutive vertices of a cycle in
$G_1$ are in variable gadgets.  This enforces the first property.

The choice of any black point in a construct of type $I$ or $II$
forces the choice of black points in this as well as in all the other
constructs of type $I$ and $II$ inside a variable gadget.  The same is
true for gray points.  In Figure~\ref{vble_gadget_points}, if the gray
point is chosen in the reference pair, gray arrows show the
implications that force the choice of gray points in all the type $I$
and $II$ constructs.  The function of the type $III$ construct in the
variable gadget is to allow this propagation.  Thus, the second
property holds; the variable gadget can be internally connected if and
only if the truth value of the reference pair agrees with the truth
value in type $I$ and type $II$ constructs.

We describe how the type $I$ and type $II$ constructs are used to
connect to clause gadgets above and below the variable gadget.
Suppose that the variable associated with this gadget is $x$.  If the
literal $x$ appears in a clause gadget embedded above the variable
gadget, then the connection from the corresponding clause gadget to
this variable gadget (to be described in the next section) is made to
the top of a construct of type $I$. In order to connect the variable
gadget to the clause gadget, a black point has to be chosen in the
reference pair.  If the literal $\overline{x}$ appears in a clause
above the variable gadget, then the connection is made to the top of a
construct of type $II$ and a gray point is chosen in a reference
pair. Similarly, if the literal $x$ appears in a clause embedded below
the variable gadget, the connection from the clause gadget is made to
the bottom of a type $II$ construct and the black point is chosen in
the reference pair. If the literal $\overline{x}$ appears in a clause
embedded below the variable gadget, the connection is made to the
bottom of a type $I$ construct and the gray point is chosen in the
reference pair.

We replace horizontal line segments corresponding to variables in the embedding of the Planar 3-SAT instance by variable gadgets. Note that the width of the $I$ and $II$ constructs can be adjusted by adding horizontally arranged uncertainty pairs.
The number of occurrences of constructs of type $I$, $II$ and $III$ depends on the number of clauses containing this variable.

\subsubsection*{Linking the gadgets}
We now explain how to represent the edges of the planar graph
$H(\Phi)$, corresponding to an instance $\Phi$ of Planar 3-SAT.  In
the embedding we are considering, edges are represented by vertical
line segments.  They represent two kinds of connections: (1) between a
pair of variables, and (2) between a clause and a variable in that
clause.  Figure~\ref{connection_points} shows vertical constructs of
uncertainty pairs that (right) connect pairs of variable gadgets and
(left) clause and variable gadgets. We observe the following
properties of the two connectors: In a clause-variable connector, the
choice of black point in a clause gadget above a variable gadget
implies the choice of black point in the variable gadget. The choice
of gray point in the clause gadget below a variable gadget forces the
choice of gray point in the variable gadget. In the variable-variable
connector, for any choice of points in the two vertically extreme
uncertainty pairs, there is a path using points in the shown
uncertainty pairs that connects the two extreme uncertainty pairs.
The white uncertainty pair allows this connection. The $n$ variable
gadgets are connected using $n-1$ variable-variable connectors that
replace $E_1$ in $H(\Phi)$.

\makeatletter{}\begin{figure}[h]
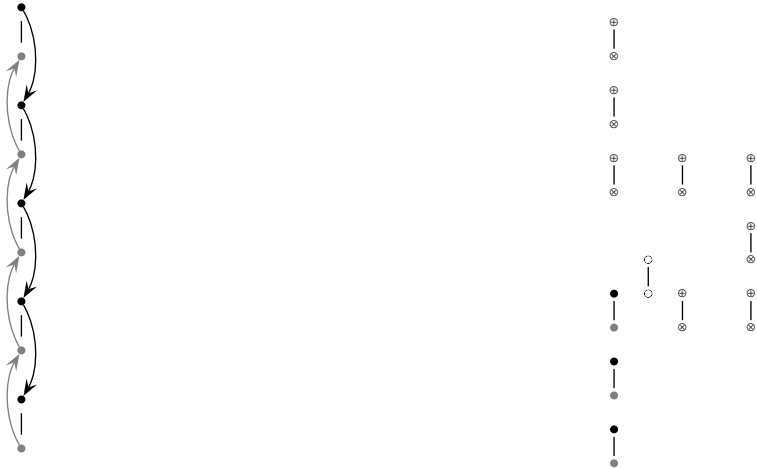


\centering
\hspace*{2cm}
\begin{tabular}{ccc}

 \psset{unit=1.3cm,arrows=-,shortput=nab,linewidth=0.5pt,arrowsize=2pt 5,labelsep=5.5pt}
 \pspicture(-1,0)(1,5)

 \psset{linecolor=gray,linewidth=0.5pt}
 \dotnode(-2,0.5){a0}
 \dotnode(-2,1.5){a1}
 \dotnode(-2,2.5){a2}
 \dotnode(-2,3.5){a3}
 \dotnode(-2,4.5){a4}

 \psset{linecolor=black,linewidth=0.5pt}
 \dotnode(-2,1){b0}
 \dotnode(-2,2){b1}
 \dotnode(-2,3){b2}
 \dotnode(-2,4){b3}
 \dotnode(-2,5){b4}

 \psset{linecolor=black,nodesep=.1,linestyle=dashed}
 \ncline{a0}{b0}
 \ncline{a1}{b1}
 \ncline{a2}{b2}
 \ncline{a3}{b3}
 \ncline{a4}{b4}

 \psset{linecolor=gray,nodesep=0,arcangle=30,linestyle=solid,arrows=->}
 \ncarc{a0}{a1}
 \ncarc{a1}{a2}
 \ncarc{a2}{a3}
 \ncarc{a3}{a4}

 \psset{linecolor=black,arrows=->}
 \ncarc{b1}{b0}
 \ncarc{b2}{b1}
 \ncarc{b3}{b2}
 \ncarc{b4}{b3}

 \psset{linecolor=black,arrows=-}
 \endpspicture

&

\  \hspace*{2.5cm}

&

 \psset{unit=0.45cm,arrows=-,shortput=nab,linewidth=0.5pt,arrowsize=2pt 5,labelsep=0.5pt}
 \pspicture(-1,-1)(5,14)

 \psset{linecolor=black,linewidth=0.5pt}
 \dotnode(0,1){b0}
 \dotnode(0,3){b1}
 \dotnode(0,5){b2}

 \psset{linecolor=black,dotstyle=o,linewidth=0.5pt}
 \dotnode(1,6){b3}

 \psset{linecolor=black,dotstyle=oplus,linewidth=0.5pt}
 \dotnode(2,5){b4}
 \dotnode(4,5){b5}
 \dotnode(4,7){b6}
 \dotnode(4,9){b7}
 \dotnode(2,9){b8}
 \dotnode(0,9){b9}
 \dotnode(0,11){b10}
 \dotnode(0,13){b11}

 \psset{linecolor=gray,dotstyle=*,linewidth=0.5pt}
 \dotnode(0,0){a0}
 \dotnode(0,2){a1}
 \dotnode(0,4){a2}

 \psset{linecolor=black,dotstyle=o,linewidth=0.5pt}
 \dotnode(1,5){a3}

 \psset{linecolor=black,dotstyle=otimes,linewidth=0.5pt}
 \dotnode(2,4){a4}
 \dotnode(4,4){a5}
 \dotnode(4,6){a6}
 \dotnode(4,8){a7}
 \dotnode(2,8){a8}
 \dotnode(0,8){a9}
 \dotnode(0,10){a10}
 \dotnode(0,12){a11}

 \psset{linecolor=black,dotstyle=*,nodesep=.1,linestyle=dashed}
 \ncline{a0}{b0}
 \ncline{a1}{b1}
 \ncline{a2}{b2}
 \ncline{a3}{b3}
 \ncline{a4}{b4}
 \ncline{a5}{b5}
 \ncline{a6}{b6}
 \ncline{a7}{b7}
 \ncline{a8}{b8}
 \ncline{a9}{b9}
 \ncline{a10}{b10}
 \ncline{a11}{b11}

 \endpspicture

\end{tabular}
\caption{(Left): A connector between clause and variable gadgets.
(Right): A connector between variable gadgets.}
\label{connection_points}
\end{figure}

Note that the parities of the (integer) heights of the tops of type I
and type II constructs in the variable gadget differ (see
Figure~\ref{vble_gadget_points}); we must ensure that clause gadgets
have the flexibility to accommodate this.  Indeed, this is the case;
consider again Figure~\ref{clause_gadget_points}.  If necessary,
gate-constructs can be shifted vertically by one unit to allow such
connections.  For example, the gate on the top of the gadget in
Figure~\ref{clause_gadget_points} can be shifted by moving all the
uncertainty pairs in the gray box up by one unit.  This change
preserves the properties of the clause gadget given earlier.

\subsubsection*{Correctness of the reduction}
In a line segment embedding of the planar graph $H(\Phi)$ corresponding to a Planar 3-SAT instance $\Phi$, nodes (clauses and variables) are horizontal line segments and edges are vertical line segments.  We have presented clause and variable gadgets to replace horizontal line segments and connectors to replace vertical line segments.  We argue that the connectivity graph $G_1$ of these uncertainty pairs is connected if and only if the Planar 3-SAT instance $\Phi$ is satisfiable.

If the Planar 3-SAT instance $\Phi$ is satisfiable, let us consider the assignment of truth values to the variables of the instance.  When a variable is set to true, we choose the black point in the reference pair of the corresponding variable gadget.  When it is set to \false, we choose the gray point.
Let $P$ be the set of points chosen and consider the graph $G_1$ of $P$. In $G_1$, the variable gadgets are all internally connected and connected to each other via the variable-variable connectors.  As all the clauses are satisfied by the truth assignment, each clause is connected to one or more variable gadgets through edge-disjoint paths. Therefore the graph $G_1$ is connected.

Let $P$ be any choice of points for which the corresponding graph $G_1$ is connected. In $G_1$, each clause gadget is connected to one or more variable gadgets through edge-disjoint paths. Therefore, two different variable gadgets can never be connected to each other through a clause gadget.
This implies that all the variable gadgets have to be internally connected and also connected to each other via the path in $H(\Phi)$ that is replaced by the $n-1$ variable-variable connectors. This structure of $G_1$ gives a truth assignment for each variable depending on whether the black or gray points are chosen inside the gadget. Since every clause gadget is connected to at least one variable gadget, the truth assignment satisfies every clause.  It is easy to see that this truth assignment satisfies the Planar 3-SAT instance $\Phi$.

This proves Theorem~\ref{theorem1}. \hfill$\Box$
\end{proof}

\subsection{Inapproximability for point pairs}
\label{subsec:best}
We observe that for uncertainty regions that are pairs of points,
there is no approximation algorithm, polynomial in the size of the
input, with an approximation ratio less than $\sqrt{5}/2$, unless
$P=NP$.  Indeed, we have provided problem instances where the
uncertainty regions are vertically aligned pairs of points separated
by a distance of one unit such that a Bottleneck Spanning Tree of
maximum edge length $2$ ($\alpha = 1$) can be found if and only if
$P = NP$.  But two points on the integer grid, if further apart than
distance $2$, must be at least distance $\sqrt{5}$ from one another.
Hence if we had a polynomial time approximation to the solution with a
ratio less than $\sqrt{5}/2$ then we could use the approximation to
find a Bottleneck Spanning Tree with maximum edge length not greater
than $2$, a contradiction (unless $P = NP$).

\begin{theorem} \label{theorem2}
There  is no approximation algorithm, polynomial in the size of the input, that solves BCU for point pairs with approximation ratio  less than $\sqrt{5}/2$, unless $P=NP$.
\end{theorem}

\subsection{BCU when the uncertainty regions are line segments}
\label{subsec:line}
We can also prove that the BCU problem is NP-hard for vertical unit segments on an integer grid.  To show this,  we use the same argument as in the proof of Theorem~\ref{theorem1} except that point pairs, unit distance apart, are now replaced by vertical line segments of unit length.

\begin{theorem} \label{theorem3}
It is NP-hard to find an exact solution to the BCU problem for the case in which the regions of uncertainty are vertical unit edges.
\end{theorem}

\noindent Hardness of approximation, such as for point pairs (Theorem~\ref{theorem2}), however, does not hold for line segments, because we can use edges of length arbitrarily close to $2$.

\subsection{BCU when the uncertainty regions are unit squares}
We prove an NP-hardness result for this problem using a reduction from
Planar 3-SAT. Our reduction uses techniques similar to the previous
reduction.

\begin{theorem}\label{theorem4}
It is NP-hard to find an exact solution to the BCU problem for the case where the regions of uncertainty are unit squares.\end{theorem}

\begin{proof}
  We use a reduction from the formulation and embedding $H(\Phi)$ of a Planar
  3-SAT instance $\Phi$ that is given in the proof of Theorem~\ref{theorem1}.

  We need the following terminology:
A point $p$ is \emph{$l$-connected} to a point $q$ if the (Euclidean) distance from $p$ to $q$ does not exceed $l$.  A set $S$ of points is $l$-connected if the maximum edge length of the minimum spanning tree of $S$ does not exceed $l$.

We now describe the variable and the clause gadgets as well as the connectors we use to link the variables and clauses.

\subsubsection{The variable gadget.}
For each variable, we create a gadget similar to the one shown in
Figure~\ref{fig:variablegadget}.  The variable gadget, as shown in the
figure, can be $5$-connected in two ways: by choosing the \redcolour\
points in each square, or by choosing the \bluecolour\ points in each
square.  We call the bold square in the figure the \emph{reference}
square.  The reference square can only be connected to at most one
square among the other squares in the variable gadget.  If the
\bluecolour\ point is chosen to make this connection in the reference
square, then we say that the variable associated with this gadget is
\true; if the \redcolour\ point is chosen, we say that the variable is
\false.  Furthermore, the subgraph of $G_1$ corresponding to a
variable gadget can be connected if and only if the points (black or
gray) chosen for all of its uncertainty regions are consistent.

\makeatletter{}\begin{figure}[h]
\begin{center}
 \psset{unit=0.3cm,arrows=-,shortput=nab,linewidth=0.5pt,arrowsize=2pt 5,labelsep=5.5pt}

 \pspicture(-1,-1)(38,26)

 \psgrid[gridcolor=lightgray,subgriddiv=1,gridwidth=0.001pt,gridlabels=0pt](0,0)(-1,-1)(38,26)

 \psline(13,0)(14,0)(14,1)(13,1)(13,0)
 \psline(23,0)(24,0)(24,1)(23,1)(23,0)
 \psline(4,3)(5,3)(5,4)(4,4)(4,3)
 \psline(9,3)(10,3)(10,4)(9,4)(9,3)
 \psline(27,3)(28,3)(28,4)(27,4)(27,3)
 \psline(32,3)(33,3)(33,4)(32,4)(32,3)
 \psline(0,7)(1,7)(1,8)(0,8)(0,7)
 \psline(13,7)(14,7)(14,8)(13,8)(13,7)
 \psline(18,7)(19,7)(19,8)(18,8)(18,7)
 \psline(23,7)(24,7)(24,8)(23,8)(23,7)
 \psline(36,7)(37,7)(37,8)(36,8)(36,7)
 \psline(0,12)(1,12)(1,13)(0,13)(0,12)
 \psline(36,12)(37,12)(37,13)(36,13)(36,12)
 \psline[linewidth=1.5pt](4,17)(5,17)(5,18)(4,18)(4,17)
 \psline(13,17)(14,17)(14,18)(13,18)(13,17)
 \psline(18,17)(19,17)(19,18)(18,18)(18,17)
 \psline(23,17)(24,17)(24,18)(23,18)(23,17)
 \psline(36,17)(37,17)(37,18)(36,18)(36,17)
 \psline(9,21)(10,21)(10,22)(9,22)(9,21)
 \psline(27,21)(28,21)(28,22)(27,22)(27,21)
 \psline(32,21)(33,21)(33,22)(32,22)(32,21)
 \psline(13,24)(14,24)(14,25)(13,25)(13,24)
 \psline(23,24)(24,24)(24,25)(23,25)(23,24)

 \psdots[linecolor=\bluecolour](13,1)(24,0)(4,4)(9,4)(27,4)(32,4)(1,8)(13,7)(18,7)(23,7)(36,7)(1,13)(36,12)(4,17)(14,18)(19,18)(24,18)(36,17)(10,21)(28,21)(33,21)(13,25)(24,24)
 \psdots[linecolor=\redcolour](13,0)(24,1)(5,4)(10,4)(28,4)(33,4)(1,7)(14,7)(19,7)(24,7)(36,8)(1,12)(36,13)(5,18)(13,18)(18,18)(23,18)(36,18)(9,21)(27,21)(32,21)(13,24)(24,25)

 \pnode(13,1){b1}
 \pnode(13.5,-1){b1x}
 \pnode(24,0){b2}
 \pnode(23.5,-2){b2x}
 \pnode(4,4){b3}
 \pnode(9,4){b4}
 \pnode(27,4){b5}
 \pnode(32,4){b6}
 \pnode(1,8){b7}
 \pnode(13,7){b8}
 \pnode(18,7){b9}
 \pnode(23,7){b10}
 \pnode(36,7){b11}
 \pnode(1,13){b12}
 \pnode(36,12){b15}
 \pnode(4,17){b16}
 \pnode(14,18){b17}
 \pnode(19,18){b18}
 \pnode(24,18){b19}
 \pnode(36,17){b20}
 \pnode(10,21){b21}
 \pnode(28,21){b22}
 \pnode(33,21){b23}
 \pnode(13,25){b24}
 \pnode(13.5,27){b24x}
 \pnode(24,24){b25}
 \pnode(23.5,26){b25x}

 \pnode(13,0){r1}
 \pnode(12.5,-2){r1x}
 \pnode(24,1){r2}
 \pnode(24.5,-1){r2x}
 \pnode(5,4){r3}
 \pnode(10,4){r4}
 \pnode(28,4){r5}
 \pnode(33,4){r6}
 \pnode(1,7){r7}
 \pnode(14,7){r8}
 \pnode(19,7){r9}
 \pnode(24,7){r10}
 \pnode(36,8){r11}
 \pnode(1,12){r14}
 \pnode(36,13){r15}
 \pnode(5,18){r16}
 \pnode(13,18){r17}
 \pnode(18,18){r18}
 \pnode(23,18){r19}
 \pnode(36,18){r20}
 \pnode(9,21){r21}
 \pnode(27,21){r22}
 \pnode(32,21){r23}
 \pnode(13,24){r24}
 \pnode(12.5,26){r24x}
 \pnode(24,25){r25}
 \pnode(24.5,27){r25x}

 \psset{linecolor=\bluecolour}
 \ncline{->}{b16}{b12}
 \ncarc[arcangle=30]{->}{b14}{b12}
 \ncarc[arcangle=30]{->}{b12}{b7}
 \ncline{->}{b7}{b3}
 \ncarc[arcangle=30]{->}{b3}{b4}
 \ncline{->}{b4}{b1}
 \ncline{->}{b4}{b8}
 \ncarc[arcangle=30]{->}{b8}{b9}
 \ncarc[arcangle=30]{->}{b9}{b10}
 \ncline{->}{b10}{b5}
 \ncarc[arcangle=30]{->}{b5}{b6}
 \ncline{->}{b6}{b11}
 \ncarc[arcangle=30]{->}{b11}{b15}
 \ncarc[arcangle=30]{->}{b15}{b20}
 \ncline{->}{b20}{b23}
 \ncarc[arcangle=30]{->}{b23}{b22}
 \ncline{->}{b22}{b19}
 \ncline{->}{b22}{b25}
 \ncarc[arcangle=30]{->}{b19}{b18}
 \ncarc[arcangle=30]{->}{b18}{b17}
 \ncline{->}{b17}{b21}
 \ncline{->}{b24}{b21}
 \ncline{->}{b2}{b5}
 \ncarc[arcangle=30]{->}{b24x}{b24}
 \ncarc[arcangle=30]{->}{b25}{b25x}
 \ncarc[arcangle=30]{->}{b1}{b1x}
 \ncarc[arcangle=30]{->}{b2x}{b2}

 \psset{linecolor=\redcolour}
 \ncline{->}{r16}{r21}
 \ncline{->}{r21}{r24}
 \ncline{->}{r21}{r17}
 \ncarc[arcangle=30]{->}{r17}{r18}
 \ncarc[arcangle=30]{->}{r18}{r19}
 \ncline{->}{r19}{r22}
 \ncline{->}{r25}{r22}
 \ncarc[arcangle=30]{->}{r22}{r23}
 \ncline{->}{r23}{r20}
 \ncarc[arcangle=30]{->}{r20}{r15}
 \ncarc[arcangle=30]{->}{r15}{r11}
 \ncline{->}{r11}{r6}
 \ncarc[arcangle=30]{->}{r6}{r5}
 \ncline{->}{r5}{r2}
 \ncline{->}{r5}{r10}
 \ncarc[arcangle=30]{->}{r10}{r9}
 \ncarc[arcangle=30]{->}{r9}{r8}
 \ncline{->}{r8}{r4}
 \ncline{->}{r1}{r4}
 \ncarc[arcangle=30]{->}{r4}{r3}
 \ncline{->}{r3}{r7}
 \ncarc[arcangle=30]{->}{r7}{r14}
 \ncarc[arcangle=30]{->}{r24}{r24x}
 \ncarc[arcangle=30]{->}{r25x}{r25}
 \ncarc[arcangle=30]{->}{r1x}{r1}
 \ncarc[arcangle=30]{->}{r2}{r2x}

 \endpspicture
\end{center}
\caption{An example variable gadget.}
\label{fig:variablegadget}
\end{figure}

Connections to clause gadgets (which replace clause vertices embedded
as horizontal lines) that contain this variable or its negation are
made via constructs similar to the four extreme top and bottom squares
shown in the example variable gadget of
Figure~\ref{fig:variablegadget}.  Of these, the top left and bottom
right connect to clauses containing this variable and the top right
and bottom left connect to clauses containing the negation of this
variable.  The width of the variable gadget can easily be increased
and more such constructs can be added to allow additional connections
(which replace edges embedded as vertical lines of $H(\Phi)$) to
clause gadgets above or below.

\subsubsection{The clause gadget.}

For each clause, we create the gadget shown in Figure~\ref{fig:clausegadget}.
We call the bold square in the clause gadget the \emph{core} square.  For each of the 3 literals (a variable or its negation), there is a sequence of squares in the gadget, called an \emph{arm}.

Observe that the core square can be $5$-connected to only a single arm, and once this arm is selected, the choice of points in its squares is \emph{fixed} in order for the squares to be $5$-connected.  The choice of points within the other arms' squares is \emph{free} since they do not need to connect to the core point.  These squares can be made connected to each other.

\makeatletter{}\begin{figure}[h]
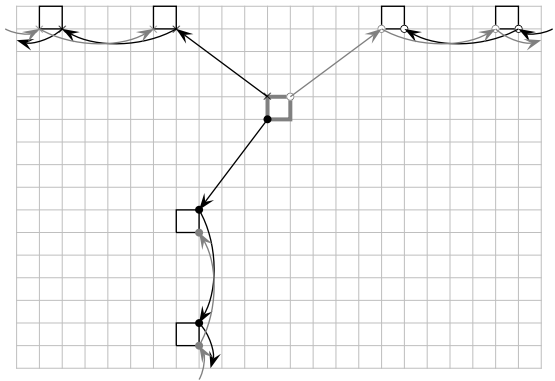

\begin{center}
 \psset{unit=0.3cm,arrows=-,shortput=nab,linewidth=0.5pt,arrowsize=2pt 5,labelsep=5.5pt}
 \pspicture(-6,3)(15,19)

 \psgrid[gridcolor=lightgray,subgriddiv=1,gridwidth=0.001pt,gridlabels=0pt](0,0)(-7,3)(16,19)

 \psline(0,4)(1,4)(1,5)(0,5)(0,4)
 \psline(0,9)(1,9)(1,10)(0,10)(0,9)
 \psline(14,18)(15,18)(15,19)(14,19)(14,18)
 \psline(9,18)(10,18)(10,19)(9,19)(9,18)
 \psline(-1,18)(0,18)(0,19)(-1,19)(-1,18)
 \psline(-6,18)(-5,18)(-5,19)(-6,19)(-6,18)
 \psline[linecolor=gray,linewidth=1.5pt](4,14)(5,14)(5,15)(4,15)(4,14)
 \psdots[linecolor=\bluecolour](1,5)(1,10)(4,14)
 \psdots[linecolor=\bluecolour,dotstyle=Bo](15,18)(10,18)
 \psdots[linecolor=\bluecolour,dotstyle=x](0,18)(-5,18)(4,15)

 \psdots[linecolor=\redcolour](1,4)(1,9)
 \psdots[linecolor=\redcolour,dotstyle=Bo](14,18)(9,18)(5,15)
 \psdots[linecolor=\redcolour,dotstyle=x](-1,18)(-6,18)

 \pnode(4,14){a1}
 \pnode(1,10){a2}
 \pnode(1,5){a3}
 \pnode(1.5,3){a4}

 \pnode(16.5,18){b0}
 \pnode(15,18){b1}
 \pnode(10,18){b2}

 \pnode(4,15){c1}
 \pnode(0,18){c2}
 \pnode(-5,18){c3}
 \pnode(-7,17.5){c4}

 \pnode(1,2.5){d0}
 \pnode(1,4){d1}
 \pnode(1,9){d2}

 \pnode(5,15){e1}
 \pnode(9,18){e2}
 \pnode(14,18){e3}
 \pnode(16,17.5){e4}

 \pnode(-7.5,18){f0}
 \pnode(-6,18){f1}
 \pnode(-1,18){f2}

 \psset{linecolor=\bluecolour}
 \ncline{->}{a1}{a2}
 \ncarc[arcangle=30]{->}{a2}{a3}
 \ncarc[arcangle=30]{->}{a3}{a4}
 \ncline{->}{c1}{c2}
 \ncarc[arcangle=30]{->}{c2}{c3}
 \ncarc[arcangle=30]{->}{c3}{c4}
 \ncarc[arcangle=30]{->}{b0}{b1}
 \ncarc[arcangle=30]{->}{b1}{b2}

 \psset{linecolor=\redcolour}
 \ncline{->}{e1}{e2}
 \ncarc[arcangle=-30]{->}{e2}{e3}
 \ncarc[arcangle=-30]{->}{e3}{e4}
 \ncarc[arcangle=-30]{->}{d0}{d1}
 \ncarc[arcangle=-30]{->}{d1}{d2}
 \ncarc[arcangle=-30]{->}{f0}{f1}
 \ncarc[arcangle=-30]{->}{f1}{f2}

 \psset{linecolor=black}

 \endpspicture
 \caption{An example clause gadget.  The \redcolour\ arrow connecting
   the core square to the top right arm represents a connection to a
   negated variable.}
\label{fig:clausegadget}
\end{center}
\end{figure}

\subsubsection{Linking the gadgets.}
Here we explain how variable gadgets can be linked to clause gadgets and how variable gadgets can be linked to each other.

In order to connect the clause gadget, shown in Figure~\ref{fig:clausegadget}, to more than one variable gadget positioned beneath it the alignment of the \redcolour\  and \bluecolour\ points in the square regions must be changed from  horizontal to vertical.  This can be accomplished using the corner gadget, shown in Figure~\ref{fig:linkgadget}.  Other useful versions of this gadget can be made by taking its horizontal and vertical mirror images. Likewise, the corner gadget enables connections to variable gadgets above the clause gadget.  To connect the lower arm of the clause gadget to a variable gadget above the clause gadget, two corner gadgets are necessary.

\makeatletter{}\begin{figure}[h]
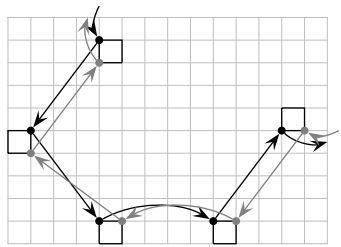

\begin{center}
 \psset{unit=0.3cm,arrows=-,shortput=nab,linewidth=0.5pt,arrowsize=2pt 5,labelsep=5.5pt}
 \pspicture(0,0)(14,10)

 \psgrid[gridcolor=lightgray,subgriddiv=1,gridwidth=0.001pt,gridlabels=0pt](0,0)(0,0)(14,10)

 \psline(0,4)(1,4)(1,5)(0,5)(0,4)
 \psline(5,8)(4,8)(4,9)(5,9)(5,8)
 \psline(4,0)(5,0)(5,1)(4,1)(4,0)
 \psline(9,0)(10,0)(10,1)(9,1)(9,0)
 \psline(12,6)(13,6)(13,5)(12,5)(12,6)
 \psset{linecolor=\bluecolour}
 \pnode(4,10.5){a0}
 \dotnode(1,5){a1}
 \dotnode(4,9){a2}
 \dotnode(4,1){a3}
 \dotnode(9,1){a4}
 \dotnode(12,5){a5}
 \pnode(14,4.5){a6}
 \ncarc[arcangle=-30]{->}{a0}{a2}
 \ncline{->}{a2}{a1}
 \ncline{->}{a1}{a3}
 \ncarc[arcangle=30]{->}{a3}{a4}
 \ncline{->}{a4}{a5}
 \ncarc[arcangle=-30]{->}{a5}{a6}

 \psset{linecolor=\redcolour}
 \pnode(3.5,10){b0}
 \dotnode(1,4){b1}
 \dotnode(4,8){b2}
 \dotnode(5,1){b3}
 \dotnode(10,1){b4}
 \dotnode(13,5){b5}
 \pnode(14.5,5){b6}
 \ncarc[arcangle=-30]{<-}{b0}{b2}
 \ncline{<-}{b2}{b1}
 \ncline{<-}{b1}{b3}
 \ncarc[arcangle=30]{<-}{b3}{b4}
 \ncline{<-}{b4}{b5}
 \ncarc[arcangle=-30]{<-}{b5}{b6}
 \psset{linecolor=black}
 \endpspicture
\caption{An example corner gadget.}
\label{fig:linkgadget}
\end{center}
\end{figure}

To complete the connection between vertex and clause gadgets, it remains to explain how to connect pairs of square uncertainty regions, one in the variable gadget and the other in an extension of one of the arms of the clause gadgets.  We would like to propagate truth assignments (choice of \redcolour\ or \bluecolour\ point in the variable gadget) consistently along such connectors.  When the distance between the points to be joined in the two squares is a multiple of $5$ and these points are vertically aligned, the connection can be made in the way analogous to that shown in Figure~\ref{connection_points}(Left).  Otherwise the construction of such connectors can be easily accomplished but is quite tedious to describe.  We leave this construction to the reader.  
To connect variable gadgets to each other (in place of the edge subset
$E_1$ of the Planar 3-SAT embedding), we add $n-1$ \emph{loose}
connections between the $n$ variable gadgets by using a sequence of
squares that allow the variable gadgets to be $5$-connected to each
other, regardless of the choice of point in the squares of the
gadget. Figure~\ref{fig:forestlink} shows an example of such a
connection.

\makeatletter{}\begin{figure}
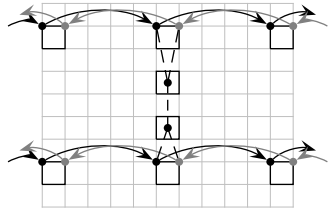

\begin{center}
 \psset{unit=0.3cm,arrows=-,shortput=nab,linewidth=0.5pt,arrowsize=2pt 5,labelsep=5.5pt}
 \pspicture(4,-1)(15,8)

 \psgrid[gridcolor=lightgray,subgriddiv=1,gridwidth=0.001pt,gridlabels=0pt](0,0)(4,-1)(15,8)

 \psline(4,0)(5,0)(5,1)(4,1)(4,0)
 \psline(9,0)(10,0)(10,1)(9,1)(9,0)
 \psline(14,0)(15,0)(15,1)(14,1)(14,0)

 \psline(4,6)(5,6)(5,7)(4,7)(4,6)
 \psline(9,6)(10,6)(10,7)(9,7)(9,6)
 \psline(14,6)(15,6)(15,7)(14,7)(14,6)

 \psline(9,2)(10,2)(10,3)(9,3)(9,2)
 \psline(9,4)(10,4)(10,5)(9,5)(9,4)

 \psdots(9.5,2.5)(9.5,4.5)
 \psset{linecolor=\bluecolour}
 \pnode(2.5,1){a2}
 \dotnode(4,1){a3}
 \dotnode(9,1){a4}
 \dotnode(14,1){a5}
 \pnode(16,1.5){a55}
 \pnode(2.5,7){a59}
 \dotnode(4,7){a6}
 \dotnode(9,7){a7}
 \dotnode(14,7){a8}
 \pnode(16,7.5){a85}

 \ncarc[arcangle=30]{->}{a2}{a3}
 \ncarc[arcangle=30]{->}{a3}{a4}
 \ncarc[arcangle=30]{->}{a4}{a5}
 \ncarc[arcangle=30]{->}{a5}{a55}
 \ncarc[arcangle=30]{->}{a59}{a6}
 \ncarc[arcangle=30]{->}{a6}{a7}
 \ncarc[arcangle=30]{->}{a7}{a8}
 \ncarc[arcangle=30]{->}{a8}{a85}

 \psset{linecolor=\redcolour}
 \pnode(3,1.5){b2}
 \dotnode(5,1){b3}
 \dotnode(10,1){b4}
 \dotnode(15,1){b5}
 \pnode(16.5,1){b55}
 \pnode(3,7.5){b59}
 \dotnode(5,7){b6}
 \dotnode(10,7){b7}
 \dotnode(15,7){b8}
 \pnode(16.5,7){b85}
 \ncarc[arcangle=30]{<-}{b2}{b3}
 \ncarc[arcangle=30]{<-}{b3}{b4}
 \ncarc[arcangle=30]{<-}{b4}{b5}
 \ncarc[arcangle=30]{<-}{b5}{b55}
 \ncarc[arcangle=30]{<-}{b59}{b6}
 \ncarc[arcangle=30]{<-}{b6}{b7}
 \ncarc[arcangle=30]{<-}{b7}{b8}
 \ncarc[arcangle=30]{<-}{b8}{b85}

 \psset{linecolor=black}
 \pnode(9.5,2.5){c1}
 \pnode(9.5,4.5){c2}

 \ncline[linestyle=dashed]{c1}{b4}
 \ncline[linestyle=dashed]{c1}{a4}
 \ncline[linestyle=dashed]{c1}{c2}
 \ncline[linestyle=dashed]{c2}{b7}
 \ncline[linestyle=dashed]{c2}{a7}

 \endpspicture
\caption{An example of a loose connection.}
\label{fig:forestlink}
\end{center}
\end{figure}

\subsubsection{Correctness of the Reduction.}

We now argue that there exists a choice of point in each of the square uncertainty regions such that the connectivity graph for $\alpha = 5/2$ is connected if and only if the corresponding Planar 3-SAT instance is satisfiable.

Suppose that there is a satisfying assignment to the Planar 3-SAT
instance.  Then, in each variable gadget we choose the \bluecolour\
corner of the reference square when that variable is \true\ and the
\redcolour\ corner when that variable is \false\ in the satisfying
assignment.
Our construction shares the property with the one in
Section~\ref{sec:point-pairs}, for point-pairs, that a variable gadget
must be internally connected for $G_1$ to be connected.  As such, we
must choose points in the other squares of the variable gadgets to be
the same colour as that of the reference square.
The choice of
\redcolour\ or \bluecolour\ point is then propagated to every clause
gadget satisfied by this assignment via the connectors.  Note that the
truth value is correctly inverted by the variable gadget when
connecting a negated variable to a clause that includes this literal.
Since all the clauses are satisfied, the core square in each clause
gadget is connected to a variable that satisfies the clause.  The arms
that do not connect to the core square are connected to their
respective variable gadgets.  At this point we have created $n$ trees,
one for each variable.  The $n-1$ connectors between variable gadgets
make $G_{5/2}$ connected.

Let $P$ be any choice of points in the square regions for which the corresponding graph $G_{5/2}$ is connected.  Since $G_{5/2}$ is connected, each clause gadget is connected to exactly one variable gadget.  Because variable gadgets can never be connected to each other via a clause gadget, each variable gadget must be internally connected.  The choice of \redcolour\ or \bluecolour\ point in the reference square of each variable gadget assigns the satisfying truth assignment to the variable associated with that variable gadget in the Planar 3-SAT instance.

By reduction from Planar 3-SAT, BCU for non-overlapping unit square uncertainty regions whose corners can be given integer coordinates is NP-hard.\hfill$\Box$
\end{proof}

\noindent Once again, hardness of approximation does not hold because we can use
edges of length arbitrarily close to $5$.

\makeatletter{}\section{An Exact Algorithm for Solving BCU for $n$ Fixed Points and $k$ Segments}\label{sec:exactAlgorithm}

We present an exact algorithm that solves BCU---for a given precision $\delta$---when the input consist of $n$ fixed points, and  the uncertainty regions are $k$ line segments of any length and orientation.  For ease of presentation, the line segments are assumed to be in general position. That is, all the line segments have different orientations. This in particular implies that no two lines are parallel. 
Our algorithm determines,  in a time that is polynomial in $n$ for any fixed $k$,  a set of point positions on the line segments, which permits a spanning tree with its longest edge being of minimum length amongst all spanning trees that connect exactly one point from each segment as well as all fixed points.

\subsubsection*{Key Ideas}
In our search for an optimum solution we focus on determining optimum solutions that satisfy additional properties that are guaranteed to exist: we seek a selection of point locations on the line segments, which supports a {\em minimum solution tree}.  A minimum solution tree is a spanning tree for segment locations and fixed points that does not just have a shortest longest edge, but also has a shortest second longest edge amongst all such solutions, and so forth. Looking for minimum solution trees, instead of optimum solutions, can decrease the search space considerably by virtue of the fact that for certain problem instances an infinite number of optimum solutions exist due to freedom of point selection on the line segments.

{\em Critical paths} give support when determining minimum solution trees. These are paths in spanning trees where all edges are of equal length and all inner path points are located on line segments.
Further, moving the location of any of the points on segments will lengthen at least one of the edges while shortening another.

Our algorithm determines point locations on all segments that support a minimum solution tree with longest tree edge of length $2\alpha$ as follows. We search for critical paths through enumerating candidates for critical paths, from longest to shortest in terms of path length. Each candidate path is tested whether or not it supports a critical path. If successful, and if the edge length of this critical path is no longer than the solution tree for the best point set found so far,  the line segments of that critical path are replaced by their corresponding point locations. We then recurse on the updated input.
Once all points on segments are determined, a greedy algorithm to determine the corresponding minimum solution tree can be applied.

\subsubsection*{Minimum Solution Trees}
To describe minimum solution trees formally, we begin by defining a
way that allows us to compare different spanning trees that correspond
to optimum solutions. We partition into equivalence classes the set of
all spanning trees taken over all fixed points and all point choices
on the $k$ segments, and we define a linear ordering on the
equivalence classes such that a minimum solution tree is a smallest
spanning tree w.r.t.~the linear ordering.

For any two selections of points on the $k$ segments, and for any two of their corresponding spanning trees, let ${\cal L}$ and ${\cal L'}$ be ordered lists of lengths of all edges in the two trees, sorted from longest to shortest. That is, ${\cal L}=(l_1,l_2,\ldots,l_{n+k-1})$ and ${\cal L'}=(l'_1,l'_2,\ldots,l'_{n+k-1})$, with $l_i\geq l_{i+1}$ and $l'_i\geq l'_{i+1}$ for all $i$.
We say that ${\cal L}$ \emph{is preferred over} ${\cal L'}$ if for a
certain $i$, $l_i<l'_i$, and $l_j=l'_j$ for all $j<i$.  If $\cal T$
and $\cal T'$ are spanning trees with edge lists $\cal L$ and
$\cal L'$, respectively, we also say $\cal T$ is preferred over
$\cal T'$ if $\cal L$ is preferred over $\cal L'$.
Note that this defines a linear ordering on lists in general, and not
only those derived from spanning tree edge-lengths.

Our algorithm seeks to choose points on segments that result in a
spanning tree such that no other spanning tree is preferred over it.
 We call such a tree a \emph{minimum solution tree} ${\cal T}$.\footnote{We remark that for lists ${\cal L}$ and ${\cal L'}$ for two different spanning trees with two different sets of points on the segments, it is possible that ${\cal L} = {\cal L'}$, so that optimum solutions that permit minimum solutions trees are in general not unique.} We call a choice of points on segments that results in a minimum solution tree  ${\cal T}$  a \emph{best point set} for  ${\cal T}$.

In a minimum solution tree ${\cal T}$, not only are longest edges as short as possible, but also the number of longest edges is minimum. In other words, a tree with a smallest number of shortest longest edges is preferred over the ones with more edges of the same length.
Further, for all $i$ the $i^{th}$  longest edge is as small as possible, and the number of edges of that length is minimum.

The above conditions imply convenient properties on the best point set w.r.t.~a minimum solution tree.
Note that, for any point $p$ on a segment in a best point set, it is impossible to {\em improve} the solution by slightly moving $p$ on its segment; in fact, any perturbation of a point must lengthen at least one of the edges
that is longest among all edges incident to $p$.
We now list the possibilities for a point $p$ on a segment in a best point set (see Figure~\ref{fig:type}) in distinguishing three different types. Given a point $p$ on a segment, we call an edge $e$ that belongs to a minimum solution tree ${\cal T}$ incident to $p$ {\em  locally longest} if no other tree edge of ${\cal T}$ incident to $p$ is longer than $e$. Then, in a minimum solution tree the possibilities for a point located on a segment  w.r.t.~its locally longest edges are as follows.

\begin{figure}[h]\centerline{\makeatletter{} \psset{unit=0.9cm,arrows=-,shortput=nab,linewidth=0.5pt,arrowsize=2pt 5,labelsep=5.5pt}
 \pspicture(0.3,-0.2)(4,3.5)

 \psline[linecolor=gray](2,0.5)(0,3.5)
 \psline[linewidth=1.5pt](1,2)(4,4)
 \uput[-30](2.5,3){segment}
 \psline[linecolor=\bluecolour](0,0)(1,2)
 \psline(2,1.5)(1,2)
 \psline(1.2,3)(1,2)
 \psdot[linewidth=1.5pt,linecolor=\redcolour](1.3,2.2)
 \psline[linecolor=\redcolour,linestyle=dashed](0,0)(1.3,2.2)
 \psdot[linewidth=1.5pt,linecolor=\bluecolour](1,2)
 \psdots(0,0)(2,1.5)(1.2,3)
\endpspicture
 
\makeatletter{} \psset{unit=0.9cm,arrows=-,shortput=nab,linewidth=0.5pt,arrowsize=2pt 5,labelsep=5.5pt}
 \pspicture(-0.6,0.3)(4,4.5)

 \psline[linecolor=gray](2,0.5)(-0.5,4.25)
 \psline[linewidth=1.5pt](-0.5,1)(2.5,3)
 \uput[-30](2.5,3){segment}
 \psline(0.5,1)(1,2)(2,1.5)
 \psline[linecolor=\bluecolour](0,3.5)(1,2)
 \psdot[linewidth=1.5pt,linecolor=\redcolour](1.3,2.2)
 \psdot[linewidth=1.5pt,linecolor=\redcolour](0.7,1.8)
 \psline[linecolor=\redcolour,linestyle=dashed](0,3.5)(1.3,2.2)
 \psline[linecolor=\redcolour,linestyle=dashed](0,3.5)(0.7,1.8)
 \psdot[linewidth=1.5pt,linecolor=\bluecolour](1,2)
 \psdots(2,1.5)(0.5,1)(0,3.5)
\endpspicture
 
\makeatletter{} \psset{unit=0.9cm,arrows=-,shortput=nab,linewidth=0.5pt,arrowsize=2pt 5,labelsep=5.5pt}
 \pspicture(-0.5,-0.2)(3.5,3.5)

 \psline[linecolor=gray](2,0.5)(0,3.5)
 \psline[linewidth=1.5pt](-0.5,1)(2.5,3)
 \uput[-30](2.5,3){segment}
 \psline[linecolor=\bluecolour](0,0)(1,2)(3,1)
 \psline(1.2,3)(1,2)
 \psdot[linewidth=1.5pt,linecolor=\redcolour](1.3,2.2)
 \psdot[linewidth=1.5pt,linecolor=\redcolour](0.7,1.8)
 \psline[linecolor=\redcolour,linestyle=dashed](0,0)(1.3,2.2)
 \psline[linecolor=\redcolour,linestyle=dashed](3,1)(0.7,1.8)
 \psdot[linewidth=1.5pt,linecolor=\bluecolour](1,2)
 \psdots(3,1)(0,0)(1.2,3)
\endpspicture
 
}\caption{Points (in \bluecolour) of Type $1$, $2$, and $3$ respectively, with longest incident edges in \bluecolour. In each case, moving the point along the segment results in a longer longest incident edge.}\label{fig:type}\end{figure}
\begin{description}
\item[Type~1] Point $p$ lies at an extremity of the segment. Then,
  one of the locally longest edge, $e$, incident to $p$  lies on
  the half plane that is delimited by a line perpendicular to the segment and
  does not contain the segment.
    We observe that moving $p$ would lengthen $e$.
 \item[Type~2] Point $p$ is on the relative interior of the segment and $e$,
one of the locally longest edges incident to $p$,
  is perpendicular to the segment. We observe that moving $p$ in any direction
  would lengthen $e$.
\item[Type~3] Point $p$ is on the relative interior of the segment but not of Type $2$.
  Then there are two locally longest edges incident to $p$ laying in different half-planes delimited by a
  line perpendicular to the segment passing through $p$. We observe that moving $p$ in any direction would increase the length of one of these two edges.
\end{description}
Notably, if we know for any point $p$ on a segment that it is of  Type $1$ or $2$ and what its locally longest incident edge $e$ is, then we can deduce $p$'s position on the segment without any knowledge of other incident edges of $p$, just by minimizing the length of $e$.
Similarly, if we know that for any point $p$ on a segment that it  is of Type $3$ and what its pair of locally longest  incident edges $e,f$ is, then we can deduce $p$'s position on the segment without any knowledge of other incident edges of $p$, just by minimizing the length of $e$ and $f$.

\subsubsection*{Critical Paths}
Let $(E_1,\ldots,E_m)$ denote a  sequence of fixed points and segments, where $E_1$ and $E_m$ are fixed points or segments, and $E_2,\ldots,$ $E_{m-1}$ are segments.
A {\em critical path supported by} $(E_1,\ldots, E_m)$ consists of
points $p_1,\ldots, p_m$, where each $p_i$ is located at a selected
position on segment $E_i$. The $p_i$s are connected by edges $e_i$
such that (1) the edges are all of identical length, and (2) no
different selection of point locations on these segments results in a
sequence where no edge is longer but some edge is strictly shorter.
We may specify that the edges are of length $\lambda$ by writing
\emph{$\lambda$-critical path}.

Critical paths are useful in constructing minimum solution trees. Our algorithm makes use of the fact that it is possible to reduce the construction of  a minimum solution tree to computing a set of critical paths.

We will show below
(1) that a sequence $(E_1, \ldots, E_m)$ supports at most one critical path (under the assumption of general position) and
(2) how to compute a critical path supported by $(E_1, \ldots,$ $E_m)$---in case of existence---for a given precision $\delta$.

We introduce terminology that will aid us for both purposes. Given sequence $(E_1,\ldots,E_m)$ and a positive number $\lambda$, let $U_i(\lambda)$ be the area around $E_i$ that can be reached from $E_1$ via $E_2, \ldots, E_{i-1}$ by edges of length at most $\lambda$. More exactly, let
\begin{itemize}
\item $U_1(\lambda)$ be the set of points in
the plane reachable from $E_1$ by an edge of length at most $\lambda$, and
\item $U_i(\lambda)$, $i>1$, be the set of points on the plane reachable
from $U_{i-1}(\lambda) \cap E_i$ by an edge of length at most $\lambda$.
\end{itemize}

Let $S_i(\lambda)$ be the set of points on the plane reachable from
$E_i$ by an edge of length exactly $\lambda$, such that
$p\in S_i(\lambda)$ implies that the path $p_1,p_2,\ldots, p_i,p$ is a
$\lambda$-critical path supported by $(E_1,E_2,\ldots, E_i,\{p\})$,
with $p_i\in E_i$ (note that $p$ need not be in one of the uncertainty
regions).  In particular, $S_i(\lambda)$ is contained, for $i>1$, in
the set of points on the plane reachable from
$S_{i-1}(\lambda) \cap E_i$ by an edge of length exactly $\lambda$.
We characterise the set $S_i(\lambda)$ more directly in Lemma 3.

\begin{figure}[h]
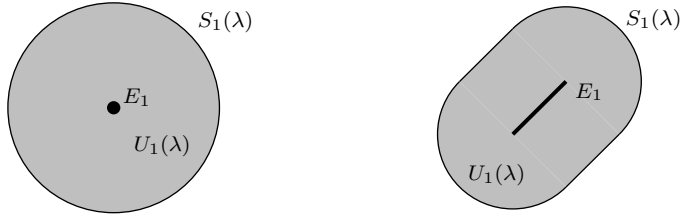

\begin{center}
\makeatletter{} \psset{unit=0.7cm,arrows=-,shortput=nab,linewidth=0.5pt,arrowsize=2pt 5,labelsep=3.5pt}
 \pspicture(-2,-2)(3,2)
 \pscircle[fillstyle=solid,fillcolor=lightgray](0,0){2}
 \psdot[linewidth=1.5pt](0,0)
 \uput[30](0,0){$E_1$}
 \uput[-30](0.2,-0.5){$U_1(\lambda)$}
 \uput[30](1.41,1.41){$S_1(\lambda)$}
\endpspicture
 
\hspace{2cm}
\makeatletter{} \psset{unit=0.7cm,arrows=-,shortput=nab,linewidth=0.5pt,arrowsize=2pt 5,labelsep=3.5pt}
 \pspicture(-1.5,-1.5)(3,2.5)
 \psarc[fillstyle=solid,fillcolor=lightgray](0,0){1.4142}{135}{315}
 \psarc[fillstyle=solid,fillcolor=lightgray](1,1){1.4142}{-45}{135}
 \psline[linestyle=none,fillstyle=solid,fillcolor=lightgray](1,-1)(2,0)(0,2)(-1,1)
 \psline(1,-1)(2,0)
 \psline(0,2)(-1,1)
 \psline[linewidth=1.5pt](0,0)(1,1)
 \uput[-30](1,1){$E_1$}
 \uput[-90](-0.3,-0.4){$U_1(\lambda)$}
 \uput[30](1.95,1.95){$S_1(\lambda)$}
\endpspicture
 
\end{center}
\caption{Examples of $U_1(\lambda)$ and $S_1(\lambda)$ for the cases that $E_1$ is a fixed point (left)
and a segment (right).}
\label{fig:e1}
\hspace{0.05cm}
\end{figure}

We study the properties of $U_i(\lambda)$ and $S_i(\lambda)$. By definition, $E_1$ consists of either a single point or a segment (Figure \ref{fig:e1}).
Further, $U_1(\lambda)$ consists either of a circle of radius $\lambda$ (Figure \ref{fig:e1}) or the Minkowski sum of a circle of radius $\lambda$ and a segment, and therefore also $U_1(\lambda) \cap E_2$---if not empty---consists of either a point or a subsegment of $E_2$.

In general we can deduce inductively the following lemma.

\begin{lemma}\label{lem:basic}
For $1< i \leq m$, if $U_{i-1}(\lambda) \cap E_i \not= \emptyset$ then $U_{i-1}(\lambda) \cap E_i$ consists of either  a single point of $E_i$ or a subsegment of $E_i$.
\end{lemma}

Using the definitions of $U_i(\lambda)$ and $S_i(\lambda)$, we have the following observation.

\begin{lemma}\label{lem:boundary}\label{subsec:l5}
For all $i \geq 1$, $S_i(\lambda)$ $\subseteq$ boundary of $U_i(\lambda)$.
\end{lemma}

\begin{proof} For $i=1$, it follows easily from the definition that,
  $S_1(\lambda)$ is contained in the boundary of $U_1(\lambda)$.  Let
  $B$ be an open ball contained in $U_i(\lambda)$, with $p\in B$.  For
  all $p'\in B$ there is a path on points $(p_1,p_2,\ldots, p_i,p')$,
  with $p_j\in E_j$, for $1\le j\le i$, with no edge longer than
  $\lambda$.  On the other hand, there exists a $p'\in B$ and path
  $(p_1,p_2,\ldots, p_i,p')$, such that
  $\|p_ip\|< \|p_ip'\|\le \lambda$, which implies that
  $p\not\in S_i(\lambda)$.  We have proved the contrapositive. $\Box$
\end{proof}

\noindent We can conclude from the above lemma that, if $U_{i-1}(\lambda) \cap E_i = \emptyset$ then $S_{i-1}(\lambda) \cap E_i =\emptyset$.
Further, if $U_{i-1}(\lambda) \cap E_i$ consists of a single point $p$, then either $S_{i-1}(\lambda) \cap E_i = \emptyset$ or $S_{i-1}(\lambda) \cap E_i = \{p\}$.
If $U_{i-1}(\lambda) \cap E_i$ is a subsegment of $E_i$, then $S_{i-1}(\lambda) \cap E_i$ can be empty, consist of one or both extremities of the subsegment, or is identical to the complete subsegment.
 In fact, we can prove the following lemma that describes $S_i(\lambda)$ more directly.

\begin{lemma}\label{subsec:l6}
\begin{enumerate}
    \item The set $S_1(\lambda)$ is the boundary of $U_1(\lambda)$.
    \item For all $i>1$, the set $S_i(\lambda)$ is the intersection of the boundary of $U_i(\lambda)$ with the Minkowski sum of a circle of radius $\lambda $ and $S_{i-1}(\lambda)\cap E_i$.
  \end{enumerate}
\end{lemma}

\begin{proof}\
\begin{enumerate}
  \item Follows from the definition of $S_1(\lambda)$ and ${U_1(\lambda)}$.
 \item
  $\subseteq$: We make two observations. (1) By definition, $S_i(\lambda) \subseteq \{$points at distance exactly $\lambda$ from $S_{i-1}(\lambda) \cap E_i\}$. This set, in turn, is contained in the Minkowski sum of a circle of radius $\lambda$ and $S_{i-1}(\lambda)\cap E_i$. (2) From Lemma~\ref{lem:boundary}, we know that
  $S_i(\lambda)$ $\subseteq$ boundary of  $U_i(\lambda)$. Combining (1) and (2) gives us the result.

  $\supseteq$: Proof by Induction.  Let $p$ be any point in the
  intersection of the boundary of $U_i(\lambda)$ with the Minkowski
  sum of a circle of radius $\lambda $ and $S_{i-1}(\lambda)\cap E_i$.
  Then, there exists $q\in S_{i-1}(\lambda)\cap E_i$ at distance
  exactly $\lambda $ from $p$ such that, by
  definition,   the path $Q=(q_1,q_2, \ldots, q_{i-1},q)$ is a $\lambda$-critical
  path supported by $(E_1,E_2,\ldots, E_{i-1},\{q\})$.

  Suppose that $p\not\in S_i(\lambda)$.  This implies that
  $(E_1, E_2,\ldots, E_{i-1}, E_i, \{p\})$ does not support a
  $\lambda$-critical path.  In particular no edge of
  $(p_1, p_2, \ldots, p_{i-1}, p_i, p)$ is longer than $\lambda$, but
  some edge is strictly shorter, for some choice of points
  $p_j\in E_j$, for $1\le j \le i$. We have $\|p_ip\|\ge \lambda$,
  since $p$ is in the boundary of $U_i(\lambda)$, so by assumption,
  $\|p_ip\|=\lambda$.  If one of the edges in
  $(p_1, p_2, \ldots, p_{i-1}, p_i)$ is shorter than $\lambda$, then
  $p_i\in E_{i}(\lambda)\setminus S_{i-1}(\lambda)$, by definition of
  $S_{i=1}(\lambda)$.  On the other hand, there is exactly one point
  in $E_i$ that is distance $\lambda$ from $p$, and thus, the
  existence of $p$ and $\|p_ip\|=\lambda$ imply that
  $p_i\in S_{i-1}\cap E_{i}$.  This is a contradiction, and therefore 
      $p\in S_i(\lambda)$, as required.

  $\Box$

  \end{enumerate}
 \end{proof}

\noindent We deduce that the following cases for $S_i(\lambda)$ are possible (see Figure \ref{fig:stype}, depicting possibilities for $S_2(\lambda)$ for the case that $E_1$ is a fixed point).

\renewcommand{\redcolour}{darkgray}

\begin{figure}[h]
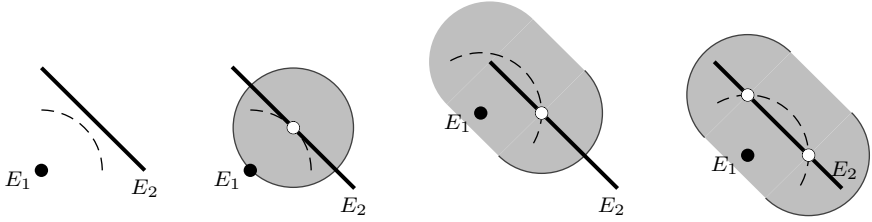

\centerline{
\makeatletter{} \psset{unit=0.8cm,arrows=-,shortput=nab,linewidth=0.5pt,arrowsize=2pt 5,labelsep=3.5pt}
 \pspicture(-0.5,-1)(2,1.7)
 \psarc[linestyle=dashed](0,0){1}{0}{90}
 \psdot[linewidth=1.5pt](0,0)
 \uput[210](0,0){$E_1$}
 \psline[linewidth=1.5pt](1.7,0)(0,1.7)
 \uput[-90](1.7,0){$E_2$}
\endpspicture 
\hspace{0.65cm}
\makeatletter{} \psset{unit=0.8cm,arrows=-,shortput=nab,linewidth=0.5pt,arrowsize=2pt 5,labelsep=3.5pt}
\pspicture(-0.5,-1)(2,1.7)
 \pscircle[fillstyle=solid,fillcolor=lightgray,linecolor=\redcolour](0.7071,0.7071){1}
 \psarc[linestyle=dashed](0,0){1}{0}{90}
 \psdot[linewidth=1.5pt](0,0)
 \psline[linewidth=1.5pt](1.7142,-0.3)(-0.3,1.7142)
 \psdot[dotstyle=o,linewidth=1.5pt,linecolor=black](0.7071,0.7071)
 \uput[210](0,0){$E_1$}
 \uput[-90](1.7,-0.3){$E_2$}
\endpspicture 
\hspace{0.65cm}
\makeatletter{} \psset{unit=0.8cm,arrows=-,shortput=nab,linewidth=0.5pt,arrowsize=2pt 5,labelsep=3.5pt}
\pspicture(-1.3,-1)(2,2.5)
 \psarc[fillstyle=solid,fillcolor=lightgray,linestyle=none](-0.3,1.8){1}{45}{225}
 \psarc[fillstyle=solid,fillcolor=lightgray,linecolor=\redcolour](0.55,0.95){1}{-135}{45}
 \psline[fillstyle=solid,fillcolor=lightgray,linestyle=none](0.4071,2.5071)(1.2571,1.6571)(-.1571,0.2429)(-1.0071,1.0929)
 \psarc[linestyle=dashed](-0.45,0.95){1}{-30}{120}
 \psdot[linewidth=1.5pt](-0.45,0.95)
 \psline[linewidth=1.5pt](1.8,-0.3)(-0.3,1.8)
 \psdot[dotstyle=o,linewidth=1.5pt,linecolor=black](0.55,0.95)
 \uput[210](-0.45,0.95){$E_1$}
 \uput[-90](1.7,-0.3){$E_2$}
\endpspicture 
\hspace{0.65cm}
\makeatletter{} \psset{unit=0.8cm,arrows=-,shortput=nab,linewidth=0.5pt,arrowsize=2pt 5,labelsep=3.5pt}
\pspicture(-0.75,-1)(2,1.7)
 \psarc[fillstyle=solid,fillcolor=lightgray,linecolor=\redcolour](0.25,1.25){1}{45}{225}
 \psarc[fillstyle=solid,fillcolor=lightgray,linecolor=\redcolour](1.25,0.25){1}{-135}{45}
 \psline[fillstyle=solid,fillcolor=lightgray,linestyle=none](0.9571,1.9571)(1.9571,0.9571)(0.5429,-0.4571)(-0.4571,0.5429)
 \psarc[linestyle=dashed](0.25,0.25){1}{-30}{120}
 \psdot[linewidth=1.5pt](0.25,0.25)
 \psline[linewidth=1.5pt](1.8,-0.3)(-0.3,1.8)
 \psdots[dotstyle=o,linewidth=1.5pt,linecolor=black](1.25,0.25)(0.25,1.25)
 \uput[210](0.25,0.25){$E_1$}
 \uput[70](1.7,-0.3){$E_2$}
\endpspicture 
}
\caption{Shapes of $S_2(\lambda)$ of Type a, b, c and d
  respectively. $S_1(\lambda)$ is indicated with a dashed line,
  $S_1(\lambda) \cap E_2$ with open dots, and $S_2(\lambda)$ with dark
  gray curves.}
\label{fig:stype}
\end{figure}

\begin{description}
  \item[Type a.] $S_{i-1}(\lambda)\cap E_i = \emptyset$ and therefore $S_i(\lambda) = \emptyset$.
  \item[Type b.] $U_{i-1}(\lambda)\cap E_i$ consists of a single point $p$ and $S_{i-1}(\lambda)\cap
    E_i = \{p\}$. Then $S_i(\lambda)$ is a circle of
    radius $\lambda$ centered around $p$.
  \item[Type c.] $U_{i-1}(\lambda)\cap E_i$ is a subsegment of $E_i$ and
    $S_{i-1}(\lambda)\cap E_i$ is a single extremity of the subsegment. In
    this case, $U_i(\lambda)$ is the Minkowski sum of the subsegment and a ball
    of radius $\lambda$, and $S_i(\lambda)$ is the half circle of radius $\lambda$ centered
    on $S_{i-1}(\lambda)\cap E_i$ on the boundary of of $U_i(\lambda)$.
   \item[Type d.] $U_{i-1}(\lambda)\cap E_i$ is a subsegment of $E_i$ and
    $S_{i-1}(\lambda)\cap E_i$ consists of both extremities of the
    subsegment. In this case, $U_i(\lambda)$ is the Minkowski sum of the
    subsegment and a ball of radius $\lambda$, and $S_i(\lambda)$ consists of both
    half circles of radius $\lambda$ each centered on a point of
    $S_{i-1}(\lambda)\cap E_i$ on the boundary of $U_i(\lambda)$.
\end{description}

\noindent For a given $(E_1,\ldots,E_m)$ and length $\lambda$ it is
therefore possible to compute successively the $U_i(\lambda)$'s and $S_i(\lambda)$'s. With this knowledge in hand, we are now ready to describe the computation of critical paths. The following lemma is crucial towards this goal.

\begin{lemma}\label{lem:critical}
A critical path exists for a sequence $(E_1,\ldots, E_m)$ if and only if there exists a $\lambda^* $ such that $U_{m-1}(\lambda^*) \cap E_m = S_{m-1}(\lambda^*) \cap E_m \neq \emptyset$. Furthermore, such a $\lambda^*$ is the smallest $\lambda$ such that $U_{m-1}(\lambda) \cap E_m \not= \emptyset$.
\end{lemma}
\begin{proof}
Figure~\ref{fig:em} illustrates the proof idea by describing types of possible outcomes for any $\lambda$ for the case of a sequence $(E_1, E_2, E_3)$ with $E_1$ a fixed point, for which $U_{2}(\lambda) \cap E_3$ is not empty.
\begin{figure}[h]
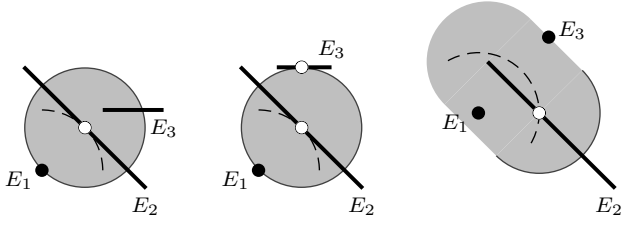

\centerline{
\makeatletter{} \psset{unit=0.8cm,arrows=-,shortput=nab,linewidth=0.5pt,arrowsize=2pt 5,labelsep=3.5pt}
\pspicture(-0,-1)(2,1.7)
 \pscircle[fillstyle=solid,fillcolor=lightgray,linecolor=\redcolour](0.7071,0.7071){1}
 \psarc[linestyle=dashed](0,0){1}{0}{90}
 \psdot[linewidth=1.5pt](0,0)
 \psline[linewidth=1.5pt](1.7142,-0.3)(-0.3,1.7142)
 \psline[linewidth=1.5pt](1,1)(2,1)
 \psdot[dotstyle=o,linewidth=1.5pt,linecolor=black](0.7071,0.7071)
 \uput[210](0,0){$E_1$}
 \uput[-90](1.7,-0.3){$E_2$}
 \uput[-90](2,1){$E_3$}
\endpspicture 
\hspace{0.75cm}
\makeatletter{} \psset{unit=0.8cm,arrows=-,shortput=nab,linewidth=0.5pt,arrowsize=2pt 5,labelsep=3.5pt}
\pspicture(-0.5,-1)(1.7,1.7)
 \pscircle[fillstyle=solid,fillcolor=lightgray,linecolor=\redcolour](0.7071,0.7071){1}
 \psarc[linestyle=dashed](0,0){1}{0}{90}
 \psdot[linewidth=1.5pt](0,0)
 \psline[linewidth=1.5pt](1.7142,-0.3)(-0.3,1.7142)
 \psline[linewidth=1.5pt](0.3,1.7071)(1.2,1.7071)
 \psdots[dotstyle=o,linewidth=1.5pt,linecolor=black](0.7071,0.7071)(0.7071,1.7071)
 \uput[210](0,0){$E_1$}
 \uput[-90](1.7,-0.3){$E_2$}
 \uput[90](1.2,1.7071){$E_3$}
\endpspicture 
\hspace{0.75cm}
\makeatletter{} \psset{unit=0.8cm,arrows=-,shortput=nab,linewidth=0.5pt,arrowsize=2pt 5,labelsep=3.5pt}
\pspicture(-1.3,-1)(1.7,2.8)
 \psarc[fillstyle=solid,fillcolor=lightgray,linestyle=none](-0.3,1.8){1}{45}{225}
 \psarc[fillstyle=solid,fillcolor=lightgray,linecolor=\redcolour](0.55,0.95){1}{-135}{45}
 \psline[fillstyle=solid,fillcolor=lightgray,linestyle=none](0.4071,2.5071)(1.2571,1.6571)(-.1571,0.2429)(-1.0071,1.0929)
 \psarc[linestyle=dashed](-0.45,0.95){1}{-30}{120}
 \psdots[linewidth=1.5pt](-0.45,0.95)(0.7071,2.2071)
 \psline[linewidth=1.5pt](1.8,-0.3)(-0.3,1.8)
 \psdot[dotstyle=o,linewidth=1.5pt,linecolor=black](0.55,0.95)
 \uput[210](-0.45,0.95){$E_1$}
 \uput[-90](1.7,-0.3){$E_2$}
 \uput[20](0.7071,2.2071){$E_3$}
\endpspicture 
}

\caption{Sequence $(E_1,E_2,E_3)$ with outcomes of Type $\alpha$,
  $\beta$, and $\gamma$ respectively for the case that
  $U_2(\lambda) \cap E_3\not= \emptyset$.  $S_1(\lambda)$ is depicted
  by dashed lines, $S_1(\lambda) \cap E_2$ and $S_2(\lambda) \cap E_3$
  by open dots, and $S_2(\lambda)$ in dark gray.}
\label{fig:em}
\end{figure}
In general for any sequence $(E_1, \dots, E_m)$ the following shapes of $S_m(\lambda)$ are possible. Recall our general position assumption that no two lines are parallel. Using Lemma~\ref{lem:basic}, either $U_{m-1}(\lambda) \cap E_m$ is a subsegment  of $E_i$ (Type $\alpha$ below) or it is a single point of $E_i$ (Type $\beta$ and $\gamma$ below).
\begin{description}
\item[Type $\alpha$.] If $E_m$ intersects the interior of $U_{m-1}(\lambda)$ then there is
  no $\lambda$-critical path.   A path with edges of length exactly $\lambda$ will not be critical as it can be improved by moving the point location on $E_3$. \item[Type $\beta$.] If $U_{m-1}(\lambda) \cap E_m = S_{m-1}(\lambda) \cap E_m = \{p\}$ then there is a $\lambda$-critical path   as moving the point location
on $E_3$ only increases the edge length.  \item[Type $\gamma$.] If $U_{m-1}(\lambda) \cap E_m$ is a single point and $S_{m-1}(\lambda) \cap E_m=\emptyset$,  there is no $\lambda$-critical path.   In this case, moving the point location on $E_2$ gives a path with shorter edge length. \end{description}
To summarize, we observe that a critical path only exists for outcomes of Type $\beta$, and that this is the only outcome for which the condition given in the lemma is satisfied. Moreover, the $\lambda^*$ in outcomes of Type $\beta$ is the smallest value of $\lambda$ for which  $U_{m-1}(\lambda) \cap E_m \not= \emptyset$ since $U_{m-1}(\lambda) \cap E_m$ is a single point and decreasing the value of $\lambda$ any further will make it empty. $\Box$
\end{proof}

Given $(E_1,\ldots, E_m)$, the following algorithm determines whether  $(E_1,\ldots, E_m)$ supports a critical path or not. If it does, the algorithm outputs the length of the edges in the critical path with a precision $\delta'$. Note that the need for a precision parameter arises only due to the algebraic nature of the problem since our computation involves binary search over real values. $\delta'$ is chosen depending on the precision bound, $\delta$, specified by the user and the input instance so that the loss in precision over all the steps of the algorithm is below the bound $\delta$. More precisely, we will choose $\delta' = \delta/k$.  We describe the reason for this choice after describing our algorithm. 

\begin{enumerate}
\item
Initialize $\lambda_{\min}$ and $\lambda_{\max}$ to be the minimum and the maximum distance between any adjacent pair $(E_i, E_{i+1})$ in the input sequence. Initialize
$\lambda$ to be $(\lambda_{\min}+\lambda_{\max})/2$.
\item
While ($\lambda_{\max}- \lambda_{\min} > \delta')$ do\\
\begin{itemize}
\item Compute $U_{m-1}(\lambda)$.
\item If $U_{m-1}(\lambda) \cap E_m = \emptyset$, then set $\lambda_{\min}$ to $\lambda$, set $\lambda$ to $(\lambda_{\min}+\lambda_{\max})/2$ and go to (2).
\item If $U_{m-1}(\lambda)\cap E_m \not= \emptyset$, then set  $\lambda_{\max}$ to $\lambda$, set $\lambda$ to $(\lambda_{\min}+\lambda_{\max})/2$ and go to (2).
\end{itemize}
\item
Check if $U_{m-1}(\lambda_{\max}) \cap E_m = S_{m-1}(\lambda_{\max}) \cap E_m \neq \emptyset$ up to perturbation $\delta'$ as explained below. If so, output $\lambda_{\max}$ as the edge-length of the critical path. If not, output that no critical path exists.
\end{enumerate}

We now give more details  about the last step of our algorithm. The algorithm will check if $U_{m-1}(\lambda_{\max}) \cap E_m$ is a line segment such that: (1) it intersects $S_{m-1}(\lambda_{\max})$ at two points and (2) its maximum distance from the boundary of $S_{m-1}(\lambda_{\max})$ is at most $\delta'$. If both (1) and (2) hold, there is an $\lambda^* \in [\lambda_{\min}, \lambda_{\max}]$ is of Type $\beta$ as illustrated in Figure~\ref{fig:em} and hence a critical path exists. For the two other possible scenarios (Type $\alpha$ and Type $\gamma$), both (1) and (2) will not be satisfied and the algorithm concludes that the critical path does not exist. These cases are also illustrated in Figure~\ref{fig:em}.

As a consequence of  Lemma~\ref{lem:critical}, we have the following corollary. It will aid us to convert a set of line segments for which we have found a critical path into a set of fixed points resulting in an input with fewer line segments.

\begin{corollary}
  If  $(E_1,\ldots,E_m)$ supports a critical path consisting of edges that are locally longest in a minimum solution tree, then there is a unique choice of point locations that defines the critical path. Furthermore, this choice of point locations is a part of the optimal solution.
\end{corollary}

\subsubsection*{Description of the Algorithm}

We now show how to compute a best point set
by examining all possible critical paths.

Note that if we had an oracle giving us a
$(E_1,\dots, E_m)$ that supports a critical path consisting of edges that are locally longest in
the best point set, then we could determine the choice of points on these
elements in the best point set. We could then replace the segments in
the sequence by fixed points and solve the rest of the problem
separately.
This eventually would allow us to replace all segments by
fixed points, and then to solve the problem by finding an associated MBST.
Lacking an
oracle
we determine
these sequences  \emph{by complete enumeration} of \emph{all possible sequences} $(E_1,\ldots,E_m)$, where $E_1$ and $E_m$ are fixed points or
segments, and $E_2,\ldots,E_{m-1}$ are segments.  There are $O((n+k)^2 \cdot k! \cdot k)$ such sequences.
This enumeration accounts for most of the complexity of
our algorithm.
For each sequence in the enumeration,
we check whether it supports a critical path and whether the edge-length for this path is best so far. If not, we discard this path.  Otherwise we recurse on the updated set of sequences and points.
That is, we prune these sequences in the enumeration as we find them. Once we have gone through the complete list of possible sequences, we will have found the best critical path with edges that are locally longest in a minimum solution tree.
Then we replace the segments of the sequence with fixed points defined by the critical path.
We then execute the algorithm recursively on the thus reduced instance, using edges of length no greater than the ones in the critical
path just found.
Once the instance does not contain any more segments, we connect all remaining connected components with a greedy algorithm, in polynomial time.

The main steps of our algorithm can be summarized as follows:
\begin{enumerate}
\item
Enumerate all possible sequences of the form $(E_1,\ldots,E_m)$, where $E_1$ and $E_m$ are fixed points or segments, and $E_2,\ldots,E_{m-1}$ are segments.
\item
For each such sequence, check if it supports a critical path and if so, compute the edge-length $l$ of the critical path and the choice of point locations in the path. 
\item
While the current (possibly partial) solution improves the previous best known solution, run the algorithm recursively on the reduced instance obtained by fixing the points locations 
on the critical path.  Look only for critical paths of edge-lengths at most $l$ and abort if no such path is found. 
\item
Continue until no line segments remain. Compute the MST on the resulting set of points using the greedy algorithm.  Update if the new MST is preferred over the 
previous best MST.
\end{enumerate}

We remark that it is crucial to determine critical paths with progressively
decreasing edge lengths since positions of points on segments
should only be determined using longest length edge incident on them. Furthermore note, when we replace the line segments with fixed points, our choice is correct up only to a precision  of $\delta'$. Since our algorithm is recursive, this choice of points will in turn influence the choice of points in the next round of recursion. Since our algorithm is computing distances between points, the edge lengths of the critical path computed in level $l$ of the recursion will be correct up to precision of $l\delta'$ (using the triangle inequality). Since the last level of recursion, level $k$, requires a precision of $\delta$, we will fix $k\delta' = \delta$ in our analysis.

The enumeration in our algorithm described above is superexponential in the number $k$ of segments,
which is not surprising since we have shown the problem with no fixed points is
NP-hard.  For constant $k$ the problem is, however, polynomial in the number $n$ of points, as our running time analysis will show.

The (multiple recursive) enumeration results in a search tree of size $O(((n+k)^2 \cdot k! \cdot k^2)^k)$  with a $O(k)$ running time for each node in the search tree. Thus the total time complexity is
$O(((n+k)^2 \cdot k! \cdot k^2)^k \cdot k)$.

\begin{theorem}
The BCU problem for a set of $n$ fixed points and $k$ line segments can be solved in time $O(((n+k)^2 \cdot k! \cdot k^2)^k \cdot k)$, up to any fixed precision $\delta$.
\end{theorem}

\makeatletter{}\section{Constant-Factor and Additive Approximations}\label{sec:constant_factor_and_additive_approx}

We begin by considering the Best-Case  Connectivity with Uncertainty problem (WCU).

\begin{lemma} \label{lemma:best_case_opt+1} Given a set of uncertainty regions that are unit disks $D_1, \ldots, D_n$ with centers $p_1, \ldots, p_n$, let $L$ be the largest edge of a minimum bottleneck spanning tree on $\{p_i: 1 \leq i \leq n\}$.  Then choosing locations $\ell_i = p_i$ and $\alpha = L/2$ is at worst an OPT$+ 1$ approximation to the BCU Problem.  In other words, if OPT denotes the smallest radius $\alpha$ for any choice of $\ell_i \in D_i$, then $L/2 \leq $ OPT$+ 1$.
This approximation can be computed in polynomial time.
\end{lemma}
\begin{proof} Consider the best choice of the $\{\ell_i \in D_i: 1 \leq i \leq n\}$ and an associated MBST on these $\{\ell_i: 1 \leq i \leq n\}$.  The edges of this MBST are each at most $2$ shorter than the corresponding edges of a spanning tree, $S$, on the corresponding $\{p_i: 1 \leq i \leq n\}$.  Thus the maximum length of any edge in $S$ is at most $2$ greater than the maximum length edge in the MBST on $\{\ell_i: 1 \leq i \leq n\}$, and, similarly, the maximum length, $L$, of any edge of a MBST on the $\{p_i: 1 \leq i \leq n\}$ must be at most $2$ greater than the maximum length edge in the MBST on $\{\ell_i: 1 \leq i \leq n\}$.  The result follows. $\Box$
\end{proof}

Our approximation for the BCU Problem, which we  dub the ``broadcast-from-center'' hueristic, is not necessarily a constant-factor approximation, because if one takes $n$ unit disks with non-empty intersection, then the $\ell_i$ can all be taken to equal one of the intersection points so that OPT$= 0$ while $L/2$ can be non-zero (and as big as $1$). However, we can modify our heuristic to obtain a constant-factor approximation for {\em non-overlapping} unit disks, for a result analogous to that obtained by Yang et al.~\cite{Yang07} for the case of MST with neighborhoods.

\begin{figure}[h]
\centerline{\scalebox{0.45}{\includegraphics{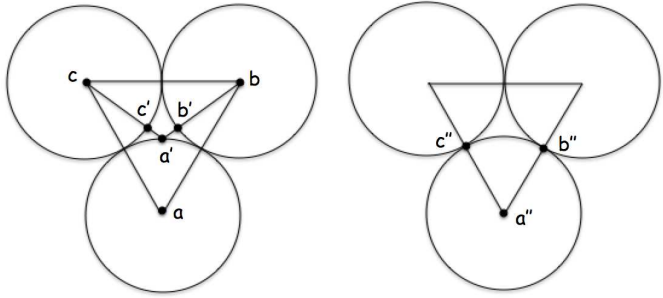}}}
\caption{The BCU Problem for three (almost) tangent unit disks.  OPT, as shown on the left is given by the choice of locations $(a',b',c')$, not quite on an equilateral triangle, with MBST cost $(\sqrt{1 + (\sqrt{3} - 1)^2} - 1)/2 \approx .12$, while the ``cinch-up'' heuristic, on the right, chooses locations $(a',b',c')$ with cost $0.5$.}
\label{3_disks}
\end{figure}

 A problem
for our heuristic, as it stands, in the case of non-overlapping disks, occurs if
we have just two disks and these two disks are within $\epsilon$ of being
tangent to one another.  As $\epsilon \rightarrow 0$ one can choose broadcast
locations $\ell_i$ increasingly close together so broadcast-from-center becomes
arbitrarily bad.  However, we can either deal with two disks as a special case,
or take the following more principled approach: begin as in
broadcast-from-center by picking the centers of all uncertainty disks, and then
find an MBST on these centers, but at the end, ``cinch-up'' any leaf nodes by
bringing the broadcast locations for these disks as close as possible to their
parent nodes.  In the case that the MBST is actually a simple path, cinch-up
twice, first at one end, then at the other.  This process ensures that we
always obtain OPT for two disks.  For three disks, we are not guaranteed to
have OPT, but because points on three unit disks cannot come arbitrarily close to
one another, the modified broadcast-from-center heuristic is a constant-factor
approximation.  See Figure \ref{3_disks}.

\section{The WCU Problem}
\label{sec:wcu}

We next consider the Worst-Case  Connectivity with Uncertainty (WCU) problem:
Find the minimum value $\alpha$ such that for {\em any} choice of points $P$, the connectivity graph $G_{\alpha}$ of $P$ is connected. In what follows we assume that the number, $n$, of points and associated uncertainty regions is at least $2$, since otherwise the problem is trivial.

We show a simple approximation algorithm for WCU that is within an additive
factor of 1 and a multiplicative factor of 2 when the uncertainty regions are
unit disks. In what follows, $D(p; \lambda)$ refers to the closed disk of radius $\lambda$ about the point $p$.

\begin{theorem} \label{lemma:worst_case_opt+1} Given a set of uncertainty regions that are unit disks $D_1, \ldots, D_n$ with centers $p_1, \ldots, p_n$, let $L$ be the largest edge of an MBST on $\{p_i: 1 \leq i \leq n\}$.  Then choosing $\alpha = L/2 + 1$ always results in the connectivity graph being connected and is at worst an OPT$+ 1$ approximation to the WCU Problem.
\end{theorem}

\begin{proof} First note that the connectivity graph given by any selection of $\ell_i \in D(p_i; 1)$ and $\{D(\ell_i;L/2 + 1): 1 \leq i \leq n\}$  is connected, because if $(p_i, p_j)$ is an edge of an MBST on $\{p_k: 1 \leq k \leq n\}$ then $D(\ell_i; L/2 + 1) \cap D(\ell_j; L/2 + 1) \neq \emptyset$.
We are thus left to show that choosing $\alpha = L/2 + 1$ is at worst an OPT$+ 1$ approximation.  But clearly we can choose $\ell_i = p_i$ for all $i$ and so the minimum $\alpha$ is $L/2$.  Hence OPT $\geq L/2$ and the theorem is established. $\Box$
\end{proof}

\begin{theorem} \label{lemma:worst_case_2*opt} Given a set of uncertainty regions that are unit disks $D_1, \ldots, D_n$ with centers $p_1, \ldots, p_n$, let $L$ be the largest edge of a MBST on $\{p_i: 1 \leq i \leq n\}$.  Then choosing $\alpha = L/2 + 1$  is at worst a factor $2$ approximation to OPT for the WCU problem.
\end{theorem}

\begin{proof} Note that OPT$+1 \leq 2$OPT as long as OPT $\geq 1$ so, by Theorem~\ref{lemma:worst_case_opt+1}, it suffices to show that OPT $\geq 1$.  If there were just one uncertainty region then our choice obviously gives OPT, hence we may assume that $n > 1$.
Let $p_j$ be a leftmost point amongst the $\{p_i: 1 \leq i \leq n\}$ and choose $\ell_j$ to be the leftmost point in $D(p_j;1)$ and for all other  $D(p_k;1)_{k \neq j}$ choose $\ell_k$ to be the rightmost point in $D(p_k;1)$.  Then $\ell_j$ is at least distance $2$ from each of the other $\ell_k$ and so we must choose $\alpha \geq 1$ to keep the connectivity graph connected.  It follows that OPT $\geq 1$ and the theorem is established. $\Box$
\end{proof}

Theorem~\ref{lemma:worst_case_2*opt} shows that the simple broadcast-from-center heuristic can be no worse than a $2$-approximation to the solution of the WCU Problem.  However, we have thus far only found examples showing that the approximation can be (asymptotically) as bad as a $\sqrt{2}$-approximation, as the next theorem asserts.

\begin{theorem} \label{lemma:worst_case_lower_bound} Given any $L > 2$, there is an instance of the WCU Problem with uncertainty regions that are unit disks
$D_1, \ldots, D_n$ with centers $p_1, \ldots, p_n$,
such that the longest edge of a MBST on the $\{p_i\}_{i=1}^n$ is $L$ and OPT is as small as $\frac{\sqrt{L^2 + 4}}{2}$, and therefore the algorithm of Theorem~\ref{lemma:worst_case_opt+1} can be as bad as a factor $\sqrt{2} - \epsilon$ approximation for arbitrarily small $\epsilon$.
\end{theorem}

\begin{proof}
We distribute an even number of unit disks with centers equally spaced along a very large (relative to the unit disks) circle $C$.  Let us call the distance between consecutive centers of disks centered along the large circle  $L$.  We will add more disks, but $L$ will remain the longest edge of a spanning tree of the disk centers.  Additionally, let us pick $\epsilon \ll L-2$.

The construction contains a large number of highly overlapping disks in addition to the disks whose centers lie along $C$.  See Figure \ref{flower} for a sketch.
\begin{figure}[h]
\centerline{\scalebox{0.40}{\includegraphics{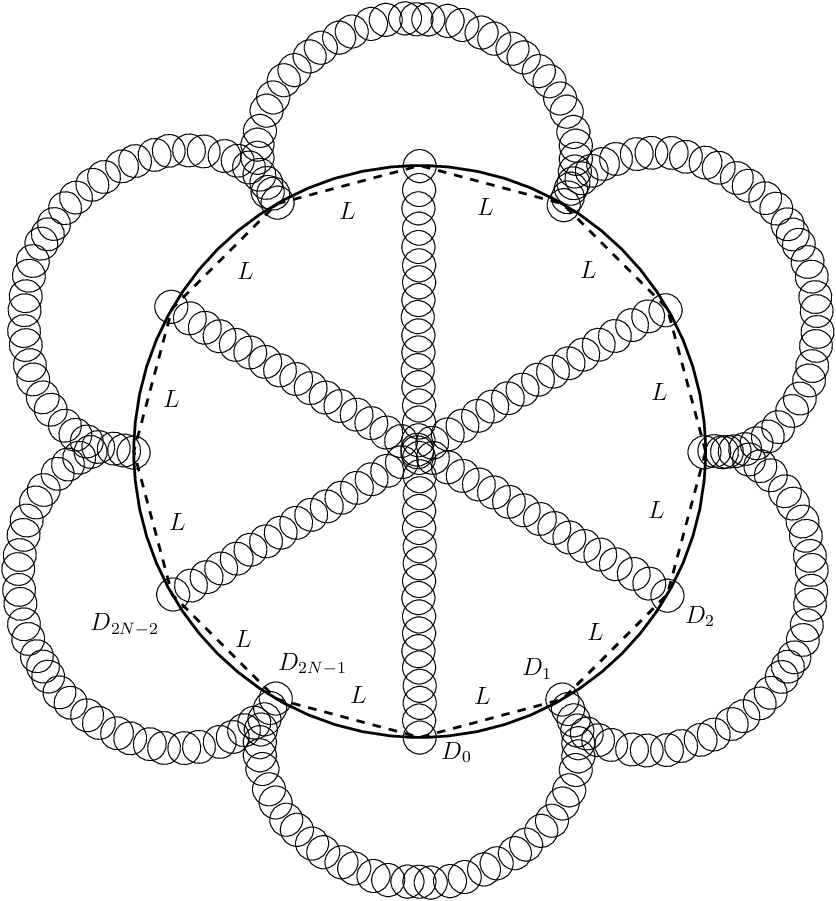}}}
\caption{The construction begins with an even number of equally spaced unit disks with centers along a very large circle $C$.}
\label{flower}
\end{figure}
The drawing is approximate in several respects.  First of all, $C$ is much, much larger than drawn, so that if the bottom of $C$ is, say, tangent to the $x$-axis, and the center of the bottom unit disk, $D_0$, has $y$-coordinate equal to $0$, then the center-points of the first unit disks to the left and right of $D_0$ along $C$, each have $y$-coordinate less than $\epsilon/3$.  In addition to the disks along $C$, there is a sequence of disks going from $D_0$ to its diametrically opposite unit disk whose centers lie along the connecting diameter. The centers of these disks are all distance $\epsilon$, one from the next, along the diameter.  If we number the $C$-centered disks in counter-clockwise order, $D_0,...,D_{2N-1}$, then we have a similar set of disks extending from $D_2, D_4,..., D_{2N-2}$.  The key observation is that we can add such diametrically centered disks in such a way that the center of disks extending from $D_j$ to the center of $C$ are each more than distance $L$ from the center of any other $D_k$ for $k \neq j$ - thus the choice of $D_1$ and $D_{2N-1}$ with $y$-coordinate less than $\epsilon/3$.  On the other hand, the odd numbered unit disks $D_1, D_3,...,D_{2N-1}$, with representative element that we shall call $D_i$, each have a set of unit disks running from $D_i$ to $D_{i+2}$ with centers each $\epsilon$ from the next, but with the disks running in almost circular patterns on the outside $C$.  An important point in this case, is that the disks start out emanating from $D_i$ along a diametric line, and then bend around so that their centers are never within $L$ of $D_{i+1}$.

We claim that for such an arrangement of unit disks, the maximum distance between locations $\ell_r \in D_r$ in a spanning tree can be as small as (and in fact slightly smaller than) $\sqrt{L^2 + 4}$, where the set $\{D_r\}$ consists not just of the disks $D_i$ with centers along $C$, but all the other unit disks depicted in Figure \ref{flower} as well.  If $D_i, D_{i+2}$ are two consecutive disks in the cyclical ordering of $C$-centered disks with $i$ odd, let $\{D_{i_k}\}$ denote the set of disks running from $D_i$ to $D_{i+2}$ outside of $C$.  Further, if $D_j, D_{j+N}$ are diametrically opposite $C$-centered disks with $j$ even, let $\{D_{j_k}\}$ denote the set of disks running diametrically between $D_j$ and $D_{j+N}$.  To verify our claim about $\{\ell_i\}$ with maximum bottleneck spanning tree edge length slightly less than $\sqrt{L^2 + 4}$, pick $\ell_i \in D_i$ for even $i$ to be the point in $D_i$ closest to the center of $C$ and $\ell_i \in D_i$ for odd $i$ to be the point in $D_i$ furthest from the center of $C$.  See Figure \ref{ells}.
\begin{figure}[h]
\centerline{\scalebox{0.50}{\includegraphics{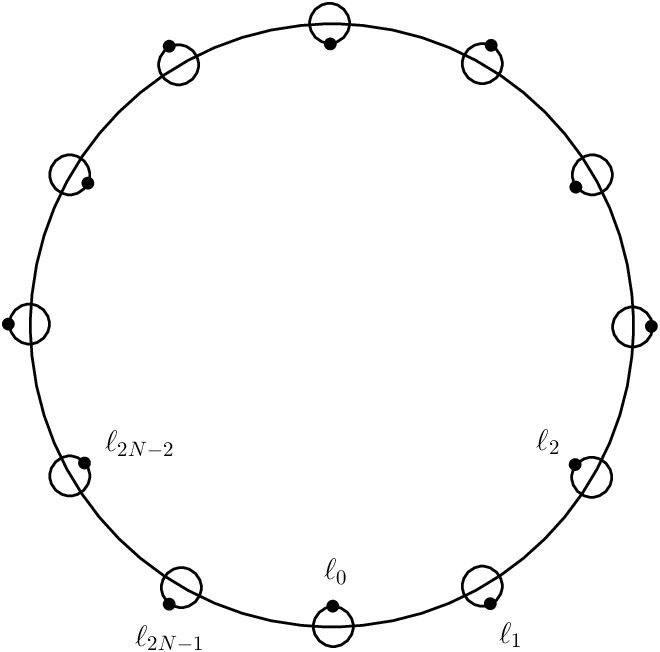}}}
\caption{The distance between $\ell_i$ and $\ell_{i+1}$ (in the cyclical ordering) is just slightly less than $\sqrt{L^2 + 4}$ since the distance between successive disk centers (which we suppose to be x-axis aligned) is $L$, and the distance between the top and bottom of the successive disks, in the y-direction, is approximately $2$.}
\label{ells}
\end{figure}
Regardless of the choice of the $\ell_{i_j} \in D_{i_j}$ it is clear that $\bigcup \{D(\ell_i;\alpha)\}~\cup~\bigcup \{D(\ell_{i_j};\alpha)\}$ is connected if $2\alpha$ is the distance between consecutive locations $\ell_i, \ell_{i+1}$ (in the cyclical ordering), and that this distance is, as claimed, just slightly less than $\sqrt{L^2 + 4}$.  Let us designate this distinguished choice of the $\ell_i \in D_i$ by $\ell_i^*$, and the associated $\alpha$ by $\alpha^*$.

For these $\{D_i\}$ and $\{D_{i_j}\}$, if there were any choice of $\{\ell_i\}, \{\ell_{i_j}\}$ making $\alpha$ any larger, then we would have to pick one of the $\ell_i$ to the left or right of the diametric line through the center of $C$ and $D_i$.
It is easy to check that the result of such a choice is that there would be some cyclically ordered pair $\ell_j, \ell_{j+1}$ whose distance $d(\ell_j, \ell_{j+1}) < d(\ell_j^*, \ell_{j+1}^*) = \alpha^*$.
But then $D(\ell_j;\alpha^*) \cup D(\ell_{j+1};\alpha^*)$ connects $\ell_j, \ell_{j+1}$ and $\bigcup \{D(\ell_{2k};\alpha^*)\}~\cup~\bigcup \{D(\ell_{(2k)_j};\alpha^*)\}$ connects the even-indexed $\ell_{2k}$ and any associated choices for $\ell_{(2k)_j}$, while $\bigcup \{D(\ell_{2k+1};\alpha^*)\}~\cup~\bigcup \{D(\ell_{(2k+1)_j};\alpha^*)\}$ connects the odd-indexed $\ell_{2k+1}$ and any associated choices for $\ell_{(2k+1)_j}$.  It follows that $\alpha \leq \alpha^*$, contrary to assumption, and so the fact that OPT can be as small as $\frac{\sqrt{L^2 + 4}}{2}$ is established.
The algorithm of Theorem~\ref{lemma:worst_case_opt+1} picked $\alpha = \frac{L}{2} + 1$, so picking $L$ sufficiently close to $2$ yields $\frac{\frac{L}{2} +1}{\frac{\sqrt{L^2 + 4}}{2}} = \frac{L+2}{\sqrt{L^2 + 4}}$ sufficiently close to
$\sqrt{2}$, completing the proof. $\Box$\\
\end{proof}

\makeatletter{}\section{Conclusions}
\label{sec:other}

A number of open problems remain.
 It would be interesting to show NP-hardness results for the BCU problem for other uncertainty regions, such as disks.
It is also possible that techniques from convex optimization could be used to design approximation algorithms for BCU for, say, line segments or squares.
We conjecture that BCU for the case of line segments is W[1]-hard and hence our exact algorithm is unlikely to be improved upon significantly.

Although we have been
able to obtain several NP-hardness results for BCU, we do not have any
complexity lower bounds for WCU which, a priori, seems harder. It is an
interesting open question to improve our approximation algorithms for both
these problems.

In conclusion, our work on connectivity problems for uncertainty regions motivated by wireless network scenarios suggests that this area provides a rich collection of problems for further investigation.

\section*{Acknowledgments}
We are grateful for two Bellairs workshops supporting this research: the 8th and 9th McGill---INRIA Workshop on Computational Geometry in 2009 and 2010. 
We also thank the anonymous reviewer for many helpful and constructive comments that greatly helped
to improve the overall presentation.

\bibliographystyle{abbrv}
\bibliography{uncertainalpha}

\end{document}